\documentclass[sigconf]{acmart}

\usepackage{subfig}
\usepackage{textcomp}
\usepackage{multirow}
\usepackage{framed}
\usepackage{amsmath}
\usepackage{listings}
\usepackage{float}

\usepackage[lined,noend,linesnumbered]{algorithm2e}

\RequirePackage[normalem]{ulem}
\RequirePackage{color}\definecolor{RED}{rgb}{1,0,0}\definecolor{BLUE}{rgb}{0,0,1}

\newcommand{\squishlist}{
 \begin{list}{$\bullet$}
  { \setlength{\itemsep}{0pt}
     \setlength{\parsep}{3pt}
     \setlength{\topsep}{3pt}
     \setlength{\partopsep}{0pt}
     \setlength{\leftmargin}{1.5em}
     \setlength{\labelwidth}{1em}
     \setlength{\labelsep}{0.5em} } }
	
\newcommand{\squishlisttwo}{
 \begin{list}{$\bullet$}
  { \setlength{\itemsep}{0pt}
     \setlength{\parsep}{0pt}
    \setlength{\topsep}{0pt}
    \setlength{\partopsep}{0pt}
    \setlength{\leftmargin}{2em}
    \setlength{\labelwidth}{1.5em}
    \setlength{\labelsep}{0.5em} } }

\newcommand{\squishend}{
  \end{list}  }

\newcommand{\eat}[1]{}

\newcommand{\captionspace}[0]{-0.4cm}

\usepackage{enumitem}
\hypersetup{
    colorlinks = true,
    allcolors = {blue}
}

\usepackage{cleveref}

\copyrightyear{2020} 
\acmYear{2020} 
\setcopyright{acmlicensed}\acmConference[HPDC '20]{Proceedings of the 29th International Symposium on High-Performance Parallel and Distributed Computing}{June 23--26, 2020}{Stockholm, Sweden}
\acmBooktitle{Proceedings of the 29th International Symposium on High-Performance Parallel and Distributed Computing (HPDC '20), June 23--26, 2020, Stockholm, Sweden}
\acmPrice{15.00}
\acmDOI{10.1145/3369583.3392675}
\acmISBN{978-1-4503-7052-3/20/06}

\begin{document}
\fancyhead{}
\pagenumbering{gobble}

\title{Cloud-scale VM Deflation for Running Interactive Applications On Transient Servers}

\author{Alexander Fuerst}
\affiliation{Indiana University}
\email{alfuerst@iu.edu}
\author{Ahmed Ali-Eldin}
\authornote{Also with Chalmers University of Technology}
\affiliation{University of Massachusetts Amherst}
\email{ahmeda@cs.umass.edu}
\author{Prashant Shenoy}
\affiliation{University of Massachusetts Amherst}
\email{shenoy@cs.umass.edu}
\author{Prateek Sharma}
\affiliation{Indiana University}
\email{prateeks@iu.edu}

\begin{abstract}
Transient computing has become popular in public cloud environments for running delay-insensitive batch and data processing applications at low cost.
Since transient cloud servers can be revoked at any time by the cloud provider, they are considered unsuitable for running interactive application such as web services.
In this paper, we present VM deflation as an alternative mechanism to server preemption for reclaiming resources from transient cloud servers under resource pressure.
Using real traces from top-tier cloud providers, we show the feasibility of using VM deflation as a resource reclamation mechanism for interactive applications in public clouds.
We show how current hypervisor mechanisms can be used to implement VM deflation and present cluster deflation policies for resource management of transient and on-demand cloud VMs.
Experimental evaluation of our deflation system on a Linux cluster shows that microservice-based applications can be deflated by up to 50\% with negligible performance overhead.
Our cluster-level deflation policies allow overcommitment levels as high as 50\%, with less than a 1\% decrease in application throughput, and can enable cloud platforms to increase revenue by 30\%.

\vspace*{-5pt}
\end{abstract}
\maketitle

\vspace*{-9pt}
\section{Introduction}
\label{sec:intro}

Transient computing is becoming commonplace in cloud environments. Today, all major cloud providers such as Amazon,  Azure,  and Google offer transient cloud servers in the form of preemptible instances that can be unilaterally revoked during periods of high server demand. Transient computing resources enable cloud providers to increase revenue by offering idle servers at significant discounts (often 7-10X cheaper) while retaining the ability to reclaim them during periods of higher demand.

While transient cloud servers have become popular due to their discounted prices, their revocable nature has meant that users  typically limit their use for running disruption-tolerant jobs such as batch or data processing tasks. They have traditionally not been used for online web services due to potential downtimes that occur when the underlying servers are revoked.

In this paper, we present virtual machine (VM) deflation as an alternative mechanism for reclaiming resources from transient cloud servers. We argue that VM deflation is more attractive than outright preemption for applications, since they continue to run, albeit more slowly, under resource pressure rather than being terminated. Deflation simplifies application design since they no longer need to implement fault tolerance approaches such as checkpointing to handle server preemptions.
Deflation also expands the classes of applications that are suitable to run on transient cloud servers---even web services can utilize such servers since downtimes from preemptions are no longer a risk; with the exception of mission critical web workloads, less critical web applications that are willing to tolerate occasional slowdowns can run on such servers at a much lower cost than on traditional cloud servers.

The notion of resource deflation was first proposed as a cascade deflation approach ~\cite{deflation-eurosys19}  that collaboratively reclaimed resources from the application, the OS, and the hypervisor. Cascade deflation requires cooperation from the OS and the application and is impractical in public clouds that treat VMs as ``black boxes.''
Instead, a hypervisor-only approach to deflation that requires no support from the application or OS is better suited to Infrastructure as a Service (IaaS) public clouds---the key focus of our work. 

By fractionally reclaiming resources from applications instead of outright preemption, VM deflation reduces the risk of downtimes for interactive applications, with a modest decrease in application performance. 
In designing and implementing  our hypervisor-only deflation approach, our paper makes the following contributions.
 
We demonstrate the feasibility of using VM deflation as a resource reclamation mechanism in public clouds using real CPU, memory, disk, and network traces from two top-tier cloud providers (Azure and Alibaba). Our analysis shows that cloud VMs running interactive applications have substantial slack and can withstand deflation of 30-50\% of their allocated resources with less than a 1\% performance impact. 

We then show how current hypervisor mechanisms such as hot-plug and throttling can be used to implement VM deflation. We also present several cluster-wide policies for VM deflation-based resource reclamation. Our policies present different tradeoffs and capabilities while attempting to minimize the performance impact of VM deflation. 

We implement a prototype of our VM deflation mechanisms and policies on a  virtualized Linux cluster and evaluate its efficacy using realistic web applications as well as other workloads. We also conduct a  trace-driven evaluation of our policies using VM-level workloads from a cloud provider.
Our results show  that:
\begin{enumerate}[leftmargin=12pt]
\vspace*{-3pt}
\item The resource utilization of cloud VMs is low, which makes deflation a viable technique for transient resources. 
\item Deflation can be implemented with hypervisor and guest-OS level overcommitment. These deflation mechanisms can reclaim large amounts of resources in a black-box manner, with minimal performance degradation. For interactive microservice based applications, even 50\% deflation results in negligible reduction in performance. 
\item Our cluster-level deflation policies make deflation an effective technique for increasing cluster overcommitment (the ratio of committed VM allocations to cluster hardware availability) by up to 50\%; nearly eliminates the risk of preemptions; and results in less than 1\% drop in application throughput. 
\end{enumerate}

\vspace*{-3pt}
The rest of this paper is structured as follows. Section \ref{sec:background} presents background on transient computing and deflation. Section \ref{sec:feasibility} presents
our feasibility analysis of VM deflation in public clouds. Section \ref{sec:mechanisms}
and \ref{sec:policies} present VM deflation mechanisms and cluster-wide deflation policies, respectively. Section \ref{sec:impl} and \ref{sec:eval} present our implementation and experimental results. Finally, Section \ref{sec:related} and \ref{sec:conclusions} present related work and our conclusions.

\vspace*{-9pt}
\section{Background}
\label{sec:background}

In this section, we provide background on transient 
cloud computing, and VM deflation.

{\noindent \bf Transient computing.}
Our work assumes a cloud data center where applications run  on traditional (``on-demand'') servers or transient servers.  Both types of servers are provisioned using virtual machines, and cloud applications run inside such VMs. 
Cloud offerings such as Amazon spot Instances~\cite{warning-time}, Google Preemptible VMs~\cite{preemptible}, and Azure batch VMs~\cite{azure-batch}  are examples of transient servers. Transient cloud servers represent surplus capacity that is offered at discounted rates but these resources can be reclaimed under resource  pressure (e.g., higher demand for on-demand servers). 

Batch-oriented applications are particularly well suited for transient computing.
Such applications tend to be both delay and disruption tolerant and can handle longer completion times.  In the event of a preemption, they can simply be restarted from the beginning or restarted from a checkpoint if the application is amenable to periodic checkpointing.  Consequently, transient cloud servers have become popular for running large batch workloads at a substantial discount over using  on-demand servers \cite{flint-eurosys16}. 

{\noindent \bf Deflation.} While current transient servers implement resource reclamation in the form of preemptions---where the VM is unilaterally revoked by the cloud provider---our work explores the use of VM deflation as an alternative approach for resource reclamation under pressure.
Although deflation frees up fewer resources than preemption (which frees up all of the VM resources), it enables applications to continue execution and eliminates application downtimes due to preempted servers \cite{deflation-eurosys19}. Our hypothesis is that occasional performance degradation, rather than termination and downtime, is more acceptable to many interactive and web applications, except the most critical ones, making transient computing feasible for a broader class of applications. 

Since modern hypervisors allow resource allocation of resident VMs to be increased or decreased dynamically, VM deflation can be realized using  current hypervisor mechanisms, such as ballooning \cite{waldspurger2002memory}, hotplugging, changing CPU shares, etc.  While any of the existing techniques can be used to implement VM deflation mechanisms, the challenge
lies in the design of judicious policies on {\em when} and {\em what} to deflate and by {\em how much}, while minimizing the impact of deflation on application performance.
We note that while VM deflation mechanisms are similar to elasticity (e.g., vertical scaling) mechanisms, our goal is to focus on cluster-wide deflation policies for resource reclamation, a different problem than elastic scaling as discussed in Section \ref{sec:related}.

\begin{figure}[t]
\vspace*{-6pt}
  \centering
  \includegraphics[width=0.3\textwidth]{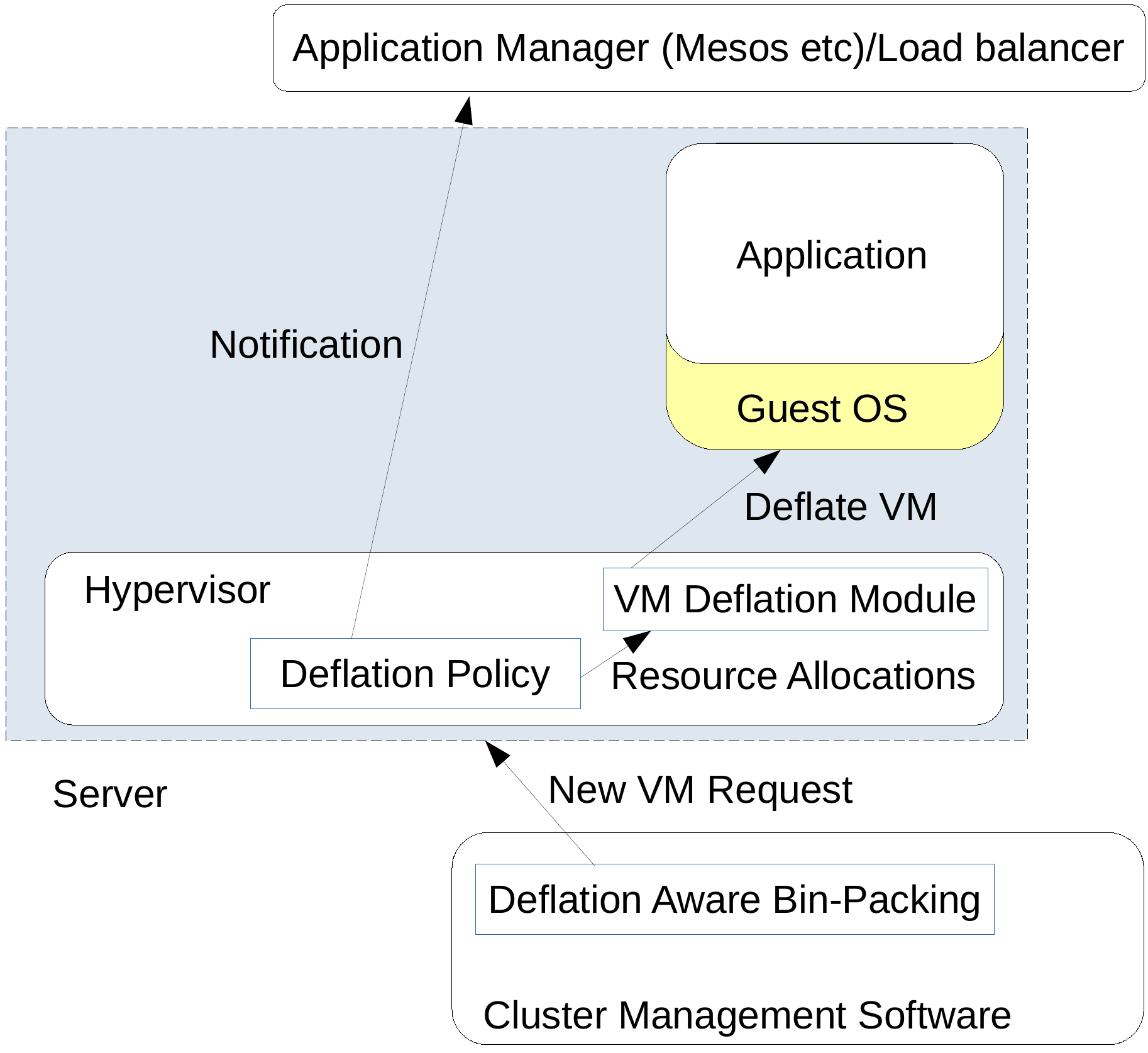}
  \vspace*{-10pt}
  \caption{Overview of our deflation system.}
  \label{fig:defl-over}
\vspace*{-10pt}
\end{figure}

Figure \ref{fig:defl-over} gives an overview of our deflation system---the cluster manager implements the global VM deflation and placement policies (Section~\ref{sec:policies}) and places new VMs onto servers.
The hypervisor implements local deflation policies (also in Section~\ref{sec:policies}), and uses VM deflation mechanisms (Section~\ref{sec:mechanisms}). 
The hypervisor also sends notifications to the application manager (such as a load balancer), which can help applications respond to deflation.

\vspace*{-5pt}
\section{Feasibility of Deflation in Public Clouds}
\label{sec:feasibility}

Before presenting our deflation techniques, we examine the efficacy and feasibility of deflating public 
cloud applications. 
We use publicly-available resource usage traces from two top-tier cloud providers, Azure~\cite{resourcecentral-sosp} and Alibaba~\cite{alibaba-trace}. 
The goal of our analysis is to understand the feasibility of deflating CPU, memory,
disk, and network allocations of real cloud applications, and specifically interactive web applications,
under time-varying workloads that they exhibit. 
We seek to answer two key research questions through our feasibility analysis: (1) How much slack is present in cloud VMs and by how much 
can these VMs be safely deflated without any performance impact?  (2) How does workload class and VM size impact the deflatability of VMs?

\begin{figure}[t]
  \centering
  \includegraphics[width=1.8in]{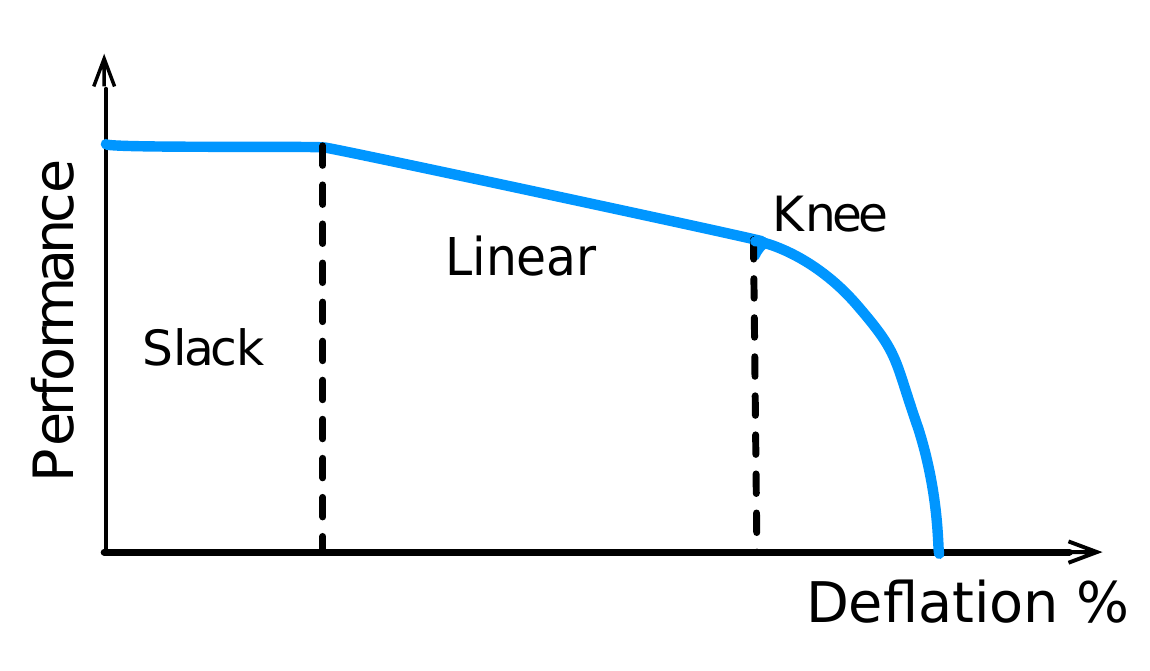}
    \vspace*{\captionspace}
  \caption{Application behavior under different levels of deflation.}
  \label{fig:deflation-model}
\vspace*{\captionspace}
\end{figure}

\subsection{Application Behavior under Deflation}
We first present an abstract model to capture the performance behavior of an application under different 
amounts of resource deflation. Figure \ref{fig:deflation-model} illustrates this behavior. We assume that an application
running inside a cloud VM will have a certain amount of slack---unused CPU and memory resources.
Reclaiming these unused resources represented by the slack will typically have negligible performance
impact on the application since they are surplus resources; the behavior in this operating region is depicted by the horizontal portion of the performance curve labelled slack in Figure \ref{fig:deflation-model}.
Once all of the slack has been reclaimed by deflating the VM, any further deflation will actually
impact performance. We assume that initially this performance impact is linear with increasing amounts
of VM deflation. For some applications, this behavior can even be sub-linear, which means that
a certain reduction in allocated resources yields proportionately less performance slowdown. For less elastic applications,
however, the impact can be super-linear.  In either case, beyond a certain point---represented by the knee of the curve---the 
performance drops precipitously, implying that allocated resources are insufficient for satisfactory performance. 

This abstract model captures the three regions with varying performance impacts on applications due to deflation.
Clearly, deflating slack is the simplest approach since it usually has little or no performance impact.
When additional resources need to be reclaimed, the deflation policy should ensure that such deflation minimizes the performance impact and does not
push application performance beyond the knee of the curve. 

\begin{figure}[t]
\centering  \includegraphics[width=0.33\textwidth]{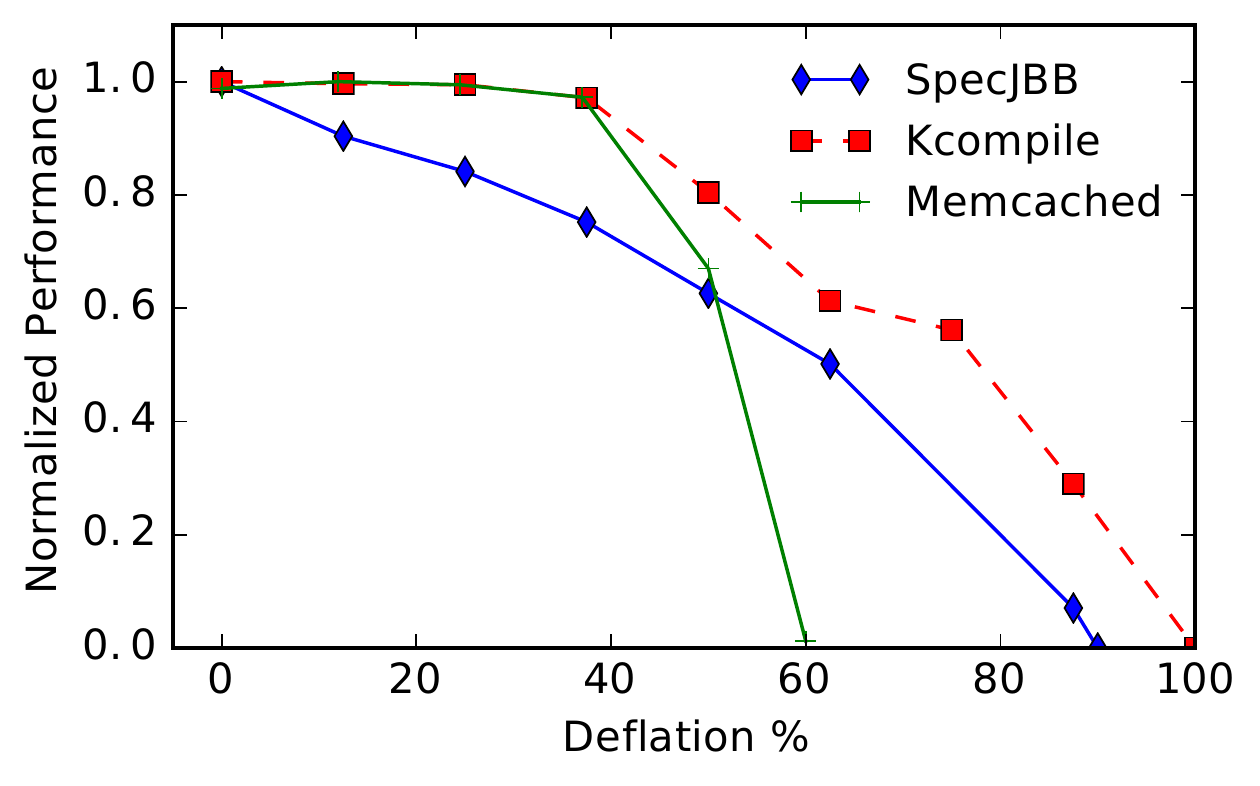}
  \vspace*{\captionspace}
  \caption{Application performance when all resources (CPU, memory, I/O) are deflated in the same proportion. }
  \label{fig:util-all}
\vspace*{-10pt}
\end{figure}

Figure \ref{fig:util-all} depicts this behavior for three different applications. As can be 
seen, different applications have different amounts of slack (with SpecJBB not exhibiting any 
slack at all in this example), and the size of the linear performance degradation region 
also varies from application to application. The figure illustrates the need to take application's
characteristics into account when reclaiming its allocated resources using deflation.

\begin{figure}[t]
  \centering
  \includegraphics[width=0.25\textwidth]{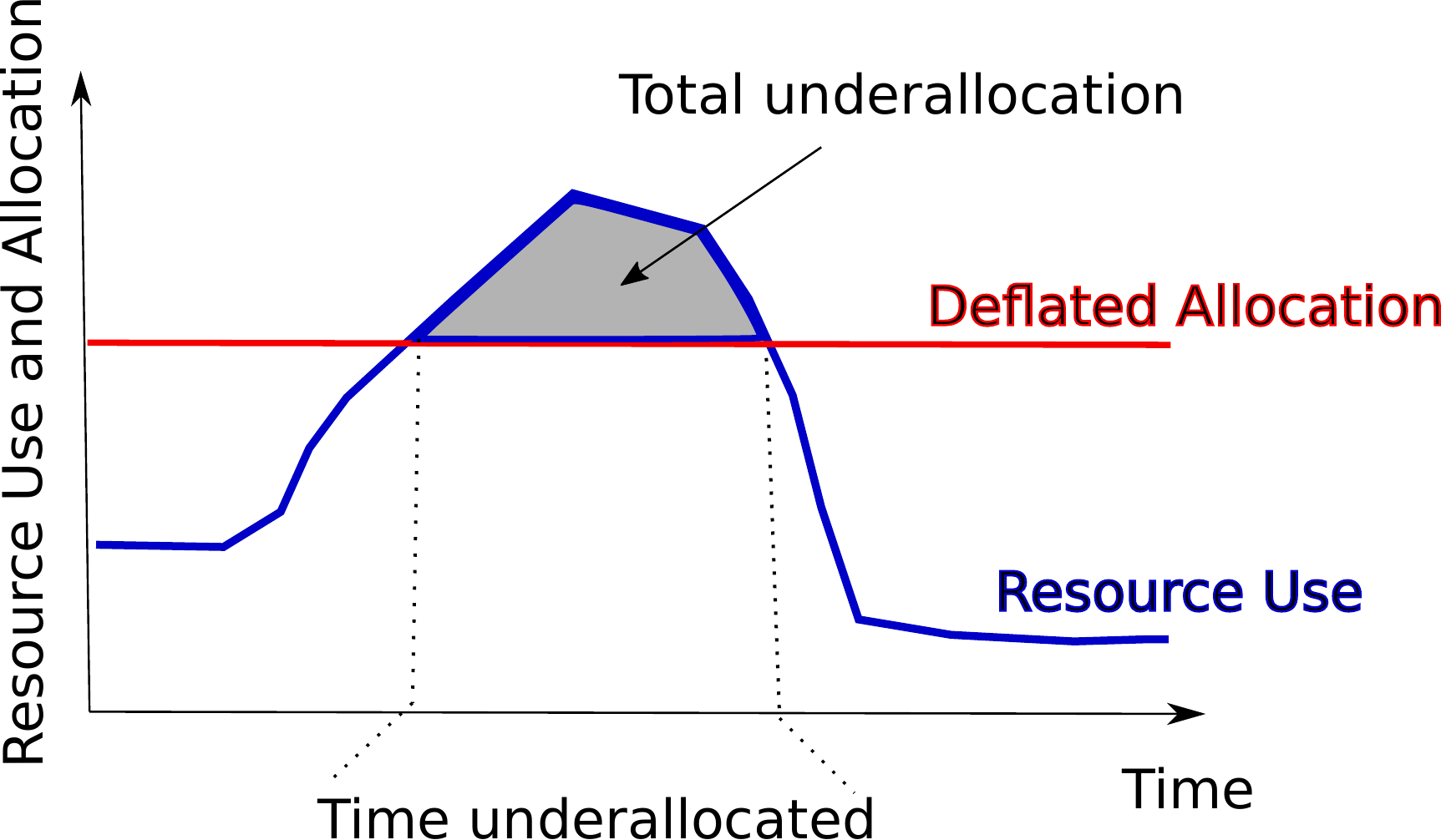}
    \vspace*{\captionspace}
  \caption{Deflation can result in underallocated resources.}
  \label{fig:underalloc}
  \vspace*{-15pt}
\end{figure}

\vspace*{-7pt}
\subsection{Usage-based Feasibility Analysis}

\subsubsection{CPU Deflation}

We analyze VM traces of CPU utilization in the Azure dataset to quantify their deflation capability. The dataset, which includes data from 2 million VMs,  provides CPU utilization time series for each VM at 5-minute granularity. Importantly for us, each VM trace is partitioned into one of three classes---interactive, delay-insensitive, and unknown---depending on the type of application resident in the VM. We analyze all three classes of VM traces but pay particular attention to interactive applications, which tend to be dominated by web-based services. 
 To analyze the impact of deflation, we assume that the CPU allocation of the VM is reduced by a certain percentage and calculate the percentage of time for which the \emph{maximum} CPU usage over each interval in the original trace exceeds this value. 
We observe that as long as the CPU utilization is below this deflated allocation, there will be no performance impact on the application.
However, during periods where the utilization exceeds the allocation under deflation (i.e., underallocation), the application will experience a slowdown.

As shown in Figure~\ref{fig:underalloc}, the resource utilization and deflation determine how much time a VM is underallocated.
The total amount of under-allocation (area of the utilization curve above the deflated allocation) is the decrease in application throughput. 
We want to quantify the slack in the VMs under different levels of deflation such that there is no performance impact on the application. 

\begin{figure}[t]
  \vspace*{\captionspace}
    \centering
    \includegraphics[width=2in]{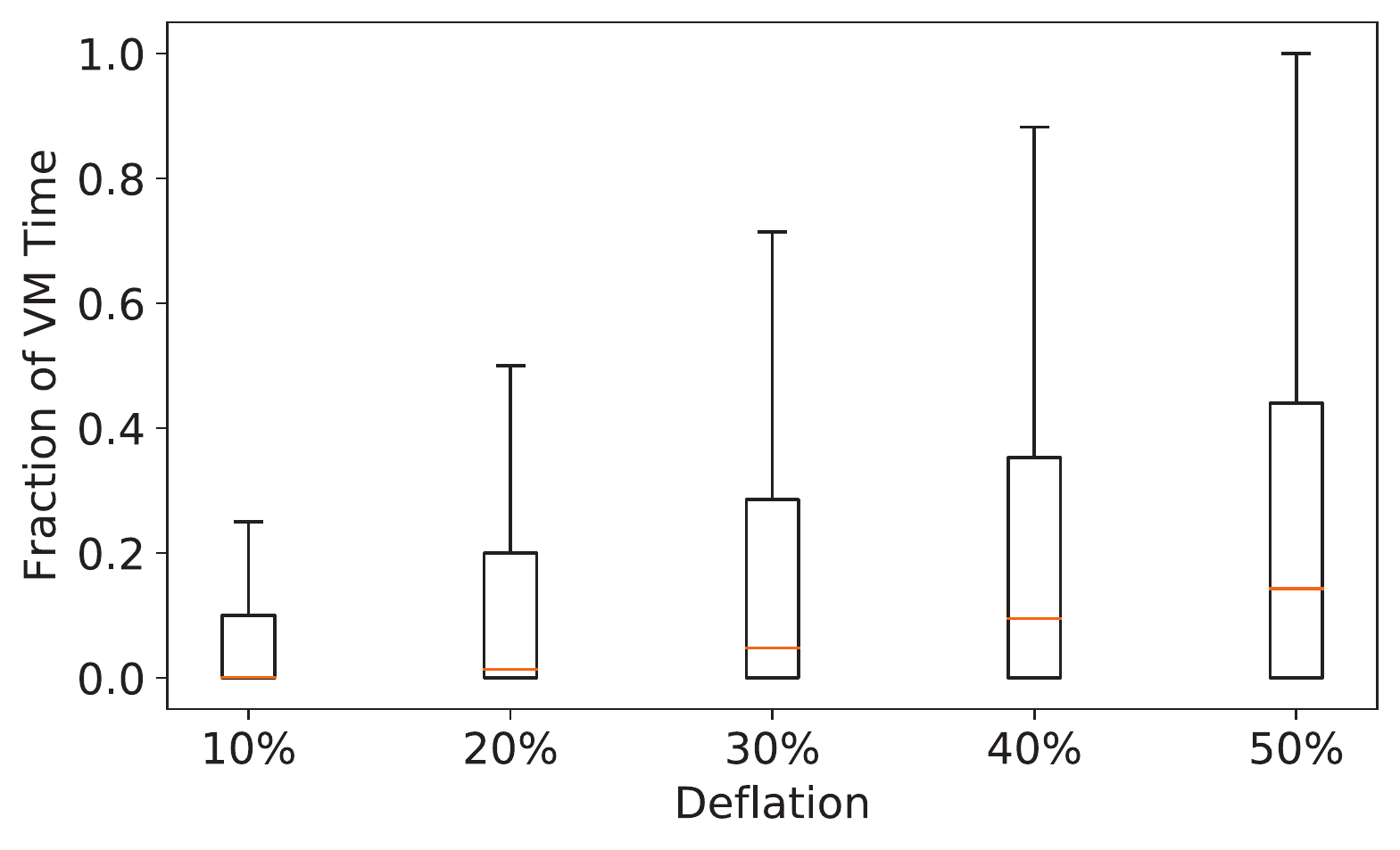}
      \vspace*{\captionspace}
      \caption{Fraction of time (i.e. probability) of CPU usage of VMs being higher than different deflation targets.}
      \label{fig:bp-thresh}
  \vspace*{-20pt}
\end{figure}

Figure~\ref{fig:bp-thresh} shows a box plot of the fraction of time spent by VMs above the deflated resource allocation (i.e., underallocated) for all 2 million VMs.
Even at high deflation levels (50\%), the median VM spends 80\% of the time below the deflated allocation.
This result indicates that even high deflation levels of as much as 50\% do not lead to significant resource bottlenecks for applications.

\begin{figure*}
\begin{minipage}[c]{0.31\linewidth}
\includegraphics[width=\linewidth]{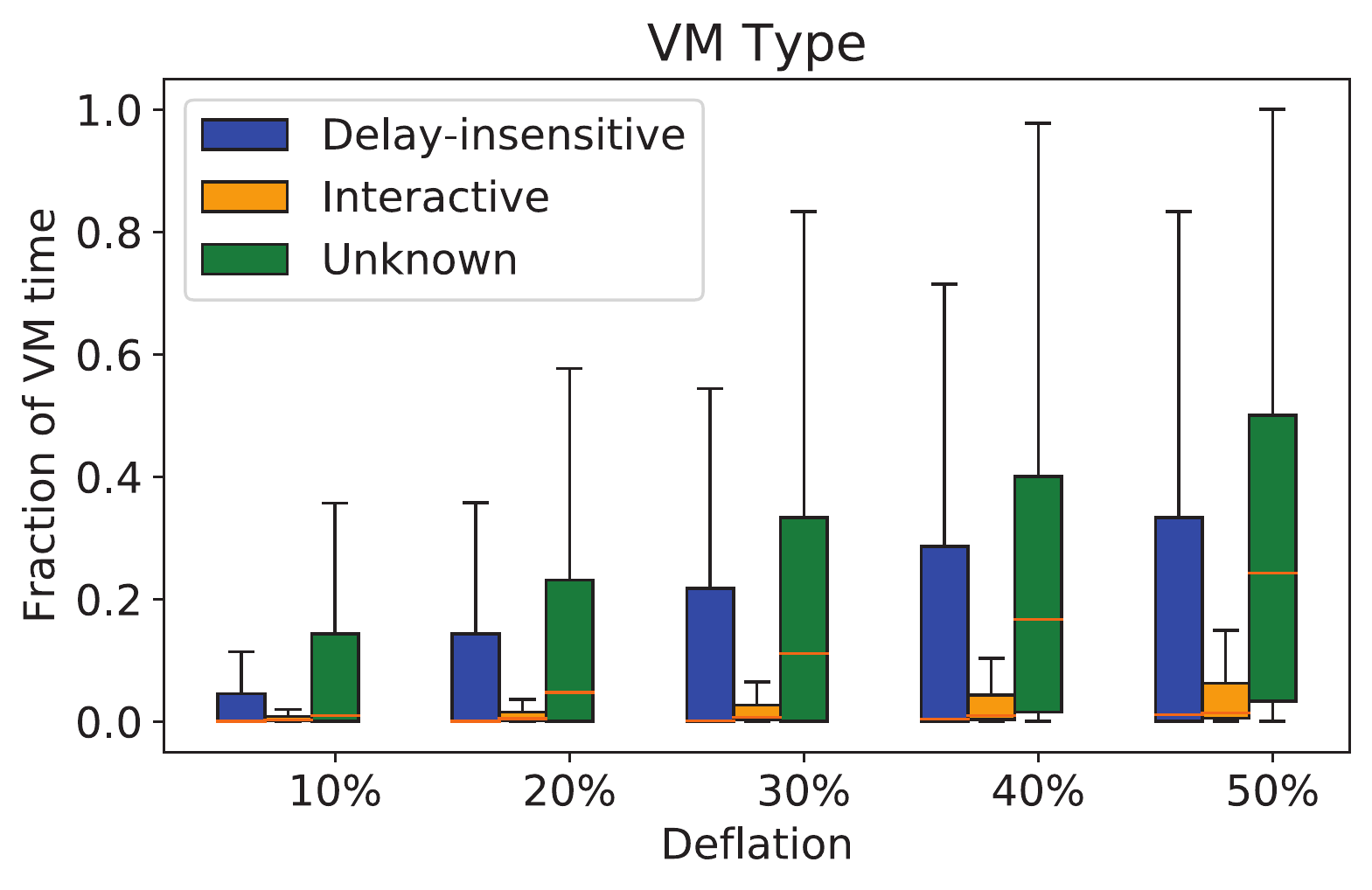}
\vspace*{-0.3in}
\caption{Fraction of time that the CPU usage of VMs is higher than different deflation targets.}
\vspace{-0.05in}
\label{fig:bp-over-thresh}
\end{minipage}
\hfill
\begin{minipage}[c]{0.31\linewidth}
\includegraphics[width=\linewidth]{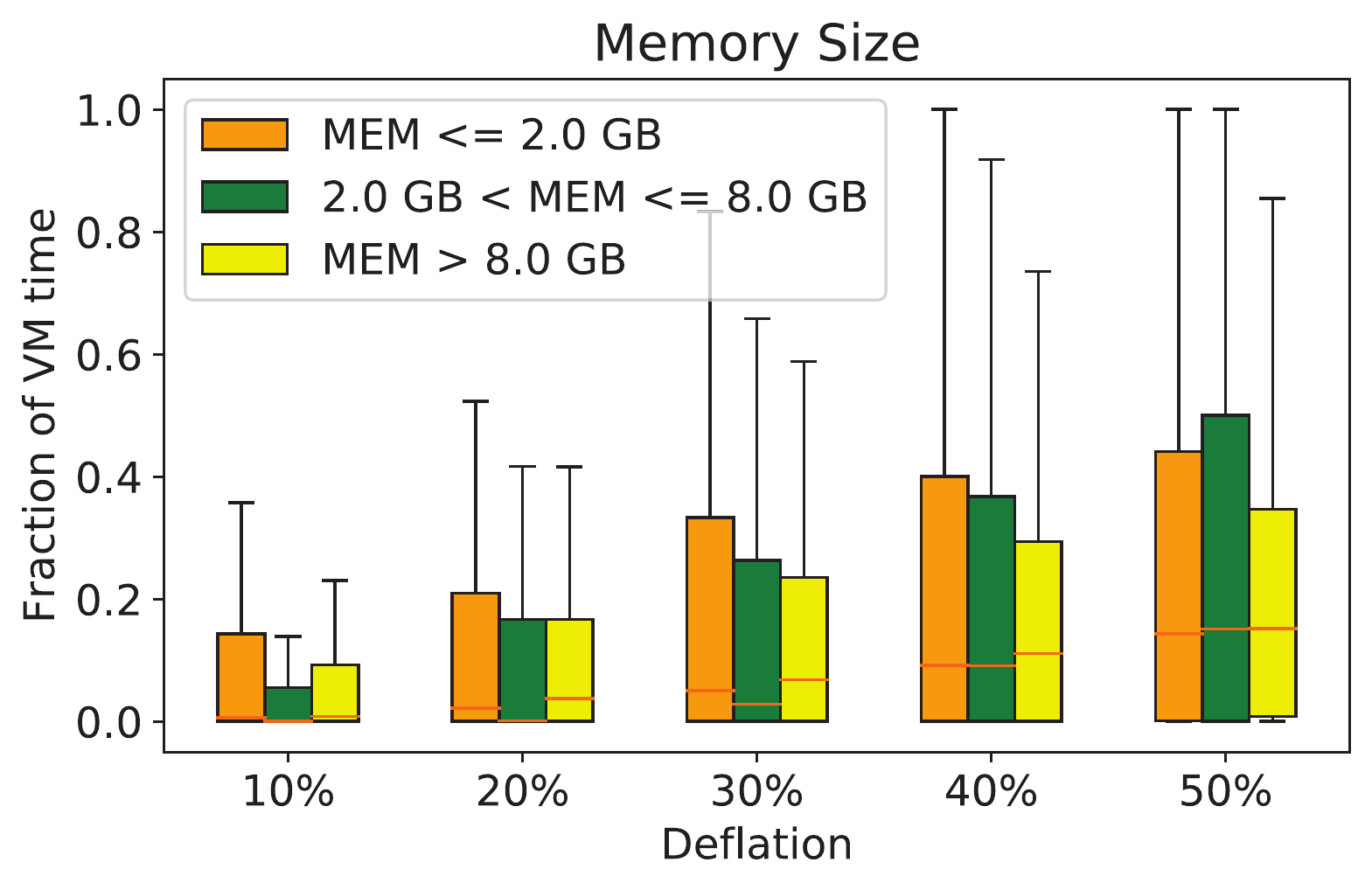}
\vspace*{-0.3in}
\caption{Breakdown of deflatability by VM memory size.}
\vspace{-0.05in}
\label{fig:bp-mem}
\end{minipage}\hfill
\begin{minipage}[c]{0.31\linewidth}
\includegraphics[width=\linewidth]{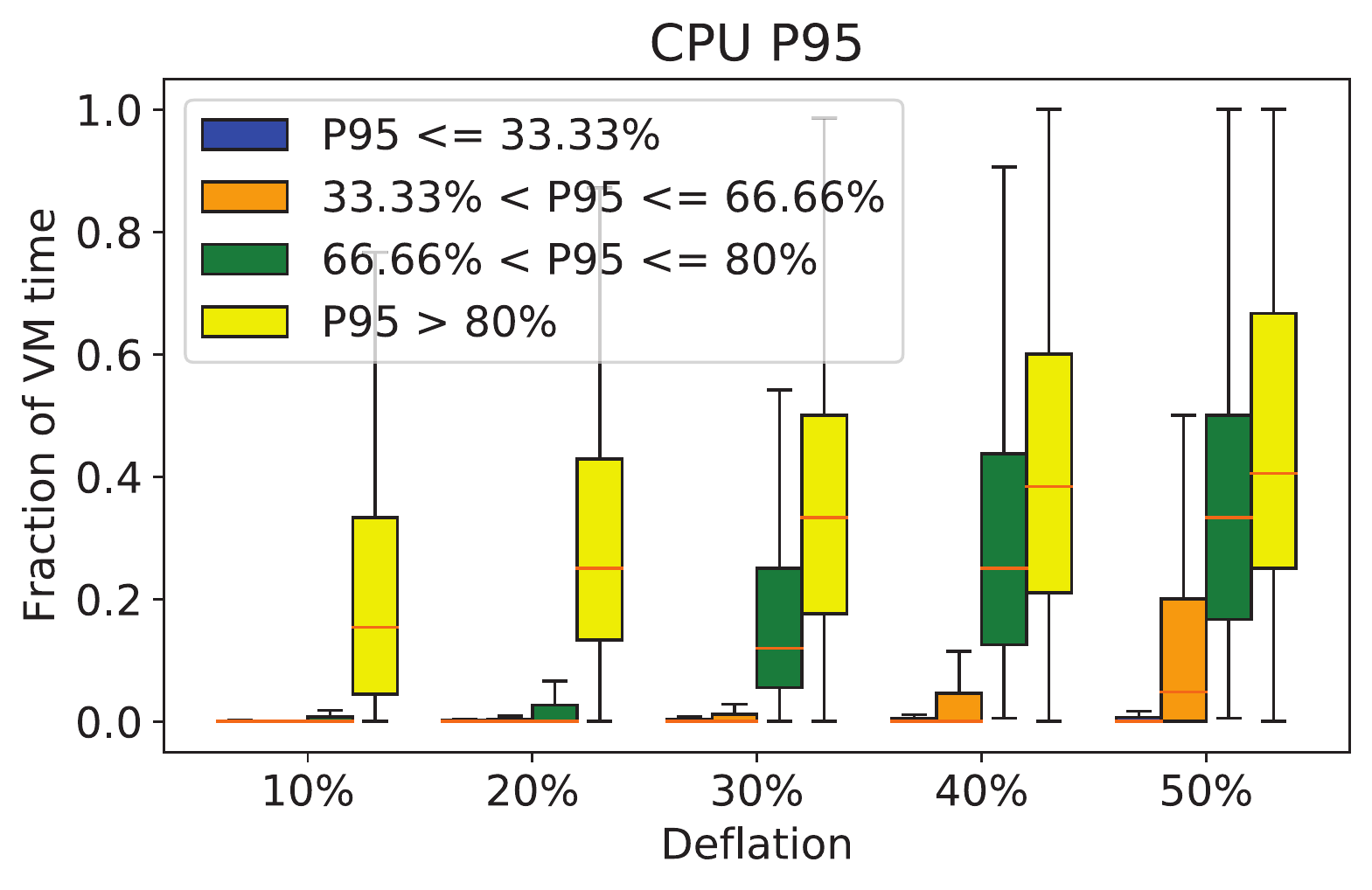}
\vspace*{-0.3in}
\caption{Breakdown of deflatability by their 95-th percentile CPU usage.}
\vspace{-0.05in}
\label{fig:bp-p95}
\end{minipage}  \vspace*{\captionspace}
\end{figure*}

Since the Azure dataset labels each VM trace with the class of application hosted by the VM, we break down the overall result in Figure~\ref{fig:bp-over-thresh} by application type. Figure~\ref{fig:bp-over-thresh} depicts a box plot of the fraction of time that VMs of different application classes exceed their deflated allocations under different levels of deflation. The figure shows that interactive applications, which include web workloads, tend to have lower overall utilization and hence more slack than delay insensitive batch workloads (presumably since they are over provisioned to handle unexpected peak loads). Consequently, interactive application VMs are more amenable to deflation of their surplus (slack) capacity. Thus, for any given deflation level, interactive VMs see significantly \emph{less} impact in terms the CPU usage exceeding the deflated allocation. The percentage of time when the interactive VMs get impacted ranges from 1\% to 15\%, as deflation percentage is varied from 10\% to 50\%. In contrast, batch jobs see 1\% to 30\% impact. This result shows that interactive applications and web workloads can be subjected to deflation just like, and perhaps more so, than delay-insensitive batch applications.

Figure~\ref{fig:bp-mem} examines whether the VM size  has an impact on its ability to be deflated. Based on the trace we partition VMs into 3 groups -- small VMs with 2 GB RAM or lower,  medium VMs with up to 8 GB RAM, and large VMs with more than 8GB RAM, and examine the percentage of time the VM CPU usage exceeds the deflated allocation within each group. The figure shows that VM size has no direct correlation to the deflatability of a VM, and all VMs see a similar performance impact under different deflation levels regardless of VM size. The result implies that VMs of all sizes are more or less equally amenable to deflation. 

Finally, Figure~\ref{fig:bp-p95} examines the deflatability of VMs for VMs with different 
peak loads. We compute the $95^{th}$ percentile of CPU usage for all VMs and partition VMs into four classes; those with low peak utilization of less than 33\%, those with moderate peak load between 33\% and 66\% peak utilization, those with higher load between 66\% and 80\% utilization and finally, the rest with high peak loads above 80\%. As shown in the figure, higher peak loads implies that VMs see greater impact when deflated since the peak will exceed the deflated allocation for longer durations. Interestingly, for deflation levels of up to 20\%, all VMs, except the ones with peak load exceeding 80\%, have enough slack to see minimal impact. The figure generally indicates that the peak load, represented by a high percentile of the utilization distribution is a coarse indicator of the ``deflatability''' of the VM; VMs with lower peak loads are more amenable to deflation.

\vspace*{-8pt}
\subsubsection{Memory and I/O Deflation}

We also analyze the memory, disk, and network deflation feasibility based on Alibaba's resource traces \cite{alibaba-analysis} \cite{alibaba-trace}, that provide a time series of resource utilization for their internal container-based interactive services.  Note that VM-based applications have a higher deflation potential because they are overprovisioned and must include additional resources for the guest OS;  thus this container-level analysis of Alibaba's cloud applications provides a very conservative (lower-bound) estimate of the actual deflation potential.

\begin{figure*}
\begin{minipage}[c]{0.23\linewidth}
\includegraphics[width=\linewidth]{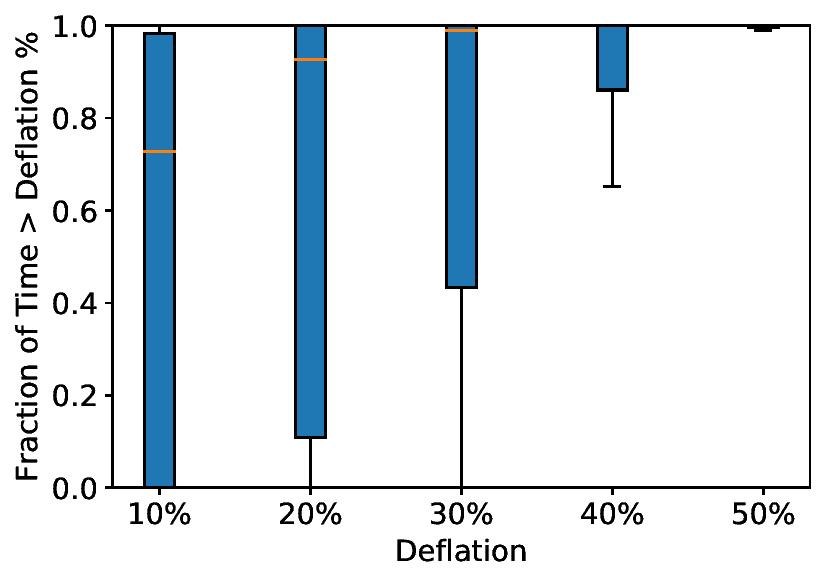}
\vspace*{-0.3in}
\caption{Memory usage of applications.}
\label{fig:ali-mem}
\end{minipage}
\hfill
\begin{minipage}[c]{0.23\linewidth}
\includegraphics[width=\linewidth]{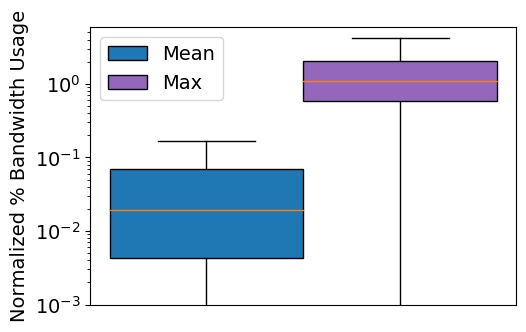}
\vspace*{-0.35in}
\caption{Memory bandwidth usage.}
\label{fig:ali-mem-band}
\end{minipage}
\hfill 
\begin{minipage}[c]{0.23\linewidth}
\includegraphics[width=\linewidth]{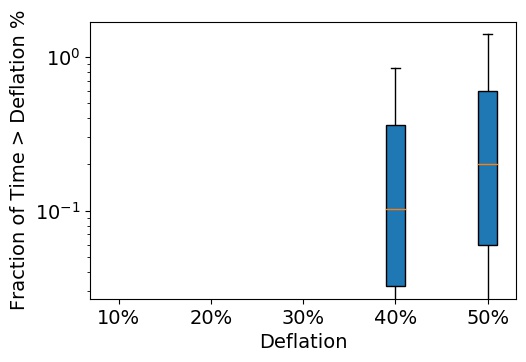}
\vspace*{-0.3in}
\caption{Disk bandwidth deflation feasibility.}
\label{fig:ali-disk}
\end{minipage}\hfill
\begin{minipage}[c]{0.23\linewidth}
\includegraphics[width=\linewidth]{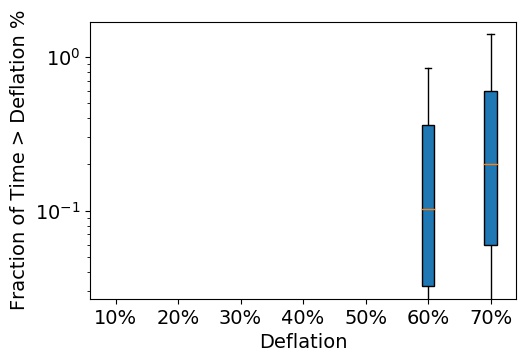}
\vspace{-0.3in}
\caption{Network bandwidth deflation feasibility.}
\label{fig:ali-net}
\end{minipage}
  \vspace*{\captionspace}
\end{figure*}

\noindent \textbf{Memory.} We analyze the memory usage of the applications under different deflation levels in Figure~\ref{fig:ali-mem}.  Interestingly, as shown, the fraction of time that the application spends above different deflation thresholds is generally high. At first glance, this might suggest that the high memory utilization leaves little slack to deflate memory (e.g., even at 10\% memory deflation,  the applications would spend more than 70\% time underallocated).

However, further analysis of the memory usage traces indicates that this is not really the  case. First, the Alibaba memory traces provide the {\em total} memory usage 
and do not provide a fine-grain breakdown of memory usage, such as such as working set size, page-cache and disk-buffer pages.  Over 90\% of the applications in Alibaba trace are JVM-based services that overallocate memory (for the heap) to reduce the garbage collection overhead. As is well known, modern applications and operating systems aggressively used unallocated RAM for purposes of caching and buffering. Hence, the total memory usage shown in Figure~\ref{fig:ali-mem} is not a true measure of deflation potential of applications.   
 
 Conventional wisdom holds that application performance will be affected when the memory is deflated below its \emph{working set} size, and deflation of other memory used for caching or garbage collection should have a lesser impact on performance.  In fact, our experiments have shown that, even when memory is deflated \emph{below} the working set size, the performance degradation, while noticeable, is not serve.  For instance, Figure~\ref{fig:util-all} shows the resilience of Memcached, a highly memory-dependent application. 
Figure~\ref{fig:jbb-mem} shows that even SpecJBB  (which is representative of the JVM-based applications that comprise the trace) can have its memory deflated by up to 30\% without significant drop in performance.

To further analyze the true memory deflation potential, we 
we use the memory-bus bandwidth used by the different applications as a proxy  metric for memory usage.
As shown in Figure~\ref{fig:ali-mem-band},
we see that the memory bandwidth usage is very low, with the mean memory bandwidth utilization across all containers being less than one-tenth of one percent, while the maximum is only 1\%.
This indicates that the applications are not reading/writing to the RAM in proportion to their memory allocations, and that the memory deflatability should be significantly higher than what is indicated by Figure~\ref{fig:ali-mem} alone.

\noindent \textbf{Disk and Network.} Finally, we examine the deflatability of disk and network bandwidth in Figures~\ref{fig:ali-disk} and \ref{fig:ali-net} using the Alibaba trace. 
We see that the usage of both I/O resources is very low.
The boxplot of application's disk bandwidth that rises above various deflation thresholds is given in Figure~\ref{fig:ali-disk}.
The percentage of time the actual disk bandwidth usage rises above various deflated allocations is low, indicating there is ample room to deflate the allocated I/O bandwidth. 
Even at a high deflation level of 50\%, containers are underallocated less than 1\% of the time. 
Network usage (sum of normalized incoming and outgoing traffic) is also low: 
in Figure~\ref{fig:ali-net} we can see that even this combined network bandwidth is not impacted by even at high (70\%) deflation levels, only suffering underallocation 1\% of their lifetime.
Below 50\% deflation, the impact is near-zero and cannot be plotted.

Our analysis shows that low-priority VMs can be shrunk to fit incoming  VMs without preemption. 
Deflation allows providers to continue offering high-priority traditional VMs, and sell unused server space for low-priority VMs that can be deflated. 
This allows consumers to still have fully-resourced VMs available for a variety of applications.
Because the average resource utilization is low, it makes sense for cloud providers to offer low-priority VMs.

\vspace*{-6pt}
\section{Deflatable Virtual Machines}
\label{sec:mechanisms}

In this section we describe how VM deflation mechanisms can be implemented using existing hypervisor mechanisms.

\vspace*{-10pt}
\subsection{VM Deflation Mechanisms}

VM deflation requires the ability to dynamically shrink  the resources allocated to the VM. 
Modern hypervisors expose interfaces to determine the current resource allocation of a VM and to dynamically modify this allocation. 
A cluster or cloud management framework can use these hypervisor APIs to implement VM deflation mechanisms.

Our system implements two classes of deflation mechanisms---{\em transparent}  mechanisms, which transparently shrink the VM's resource allocation, and {\em explicit}  mechanisms, where the deflation is performed in a manner that is visible to the guest OS, (and by extension, to the applications and the application cluster manager). 
In the former case, the guest OS and applications are unaware of the deflation and the VM simply runs ``slower'' than prior to deflation. 
In the latter case, since deflation is visible to the guest OS and/or applications, they can take explicit measures, if wanted, to deal with deflation. 
We describe each mechanism and a hybrid approach that exploits the key benefits of both approaches.

\vspace*{-7pt}
\subsection{Transparent VM Deflation}
\vspace*{-4pt}
Since hypervisors offer virtualized resources to virtual machines, they can also \emph{overcommit} these resources by multiplexing virtual resources onto physical ones. 
Transparent VM deflation is implemented using these hypervisor overcommitment mechanisms.
For example, the hypervisor allows virtual CPUs (vCPUs) of the VM to be mapped onto  dedicated physical CPU cores. 
Such an allocation can be deflated by remapping the vCPUs onto a smaller number of physical cores using the hypervisor's CPU scheduler. 
Thus the guest OS and applications inside the VM still see the same number of vCPUs, but these vCPUs run slower. 

In the case of memory, hypervisors allocate an amount of physical memory to a VM and multiplexes the VM's virtualized memory address-space onto physical memory, via two-dimensional paging. 
Memory deflation thus involves dynamically reducing the physical memory allocated to a VM. 

In the case of network, one or more logical network interfaces of a VM are mapped onto one or more physical NICs and a certain bandwidth of the physical NICs is allocated to each vNIC by the hypervisor.  
Network deflation involves reducing the physical NIC bandwidth allocated to the VM.
Finally, in the case of local disks, the I/O bandwidth allocated to each VM can be throttled.
With the above hypervisor level transparent techniques, the VM and applications are oblivious of the deflation, which is done at the hypervisor level outside of the VM.  
The VM may get scheduled at a lower frequency or have less physical memory, etc. 
Our deflation framework has been implemented in KVM and Linux using Linux's cgroups facility. 
By running KVM VMs inside of cgroups, we can control the physical resources available for the VM to use. 
For deflating CPUs, we use CPU bandwidth control by setting the CPU shares of the deflatable VM. 
The memory footprint of a deflatable VM is controlled by restricting the VM's physical memory allocation by setting the memory limit in the memory cgroup. 
Similarly for disk and network I/O, we use the respective I/O cgroups to set bandwidth limits.

\vspace*{-7pt}
\subsection{Explicit Deflation via Hotplug}
\vspace*{-4pt}
Modern virtualization environments now support the ability to explicitly hot plug (and unplug) resources from running guest operating systems. 
Explicit deflation mechanisms use these hot unplug techniques to reduce the VM's allocation in a manner that is visible to the guest OS and the applications. 
In the case of CPU, if a VM has $n$ vCPUs allocated to it, its CPU resources are reclaimed by unplugging  $k$ out of $n$ vCPUs. 
Hot plugging and unplugging requires guest OS support, since it must reschedule/rebalance processes and threads to a smaller or larger number of cores. Thus, the deflation is visible to the guest OS and applications. 
In the case of memory, we use memory unplugging to inform the OS and applications of the resource pressure, which allows them to return unused pages, shrink caches, etc. 
Explicit unplugging of NICs and disks is generally unsafe, and we rely on the transparent hypervisor-level mechanisms for these. 

Hot unplugging has a safety threshold---unplugging too many resources (e.g., too much memory) beyond this safety threshold can cause OS or application failures. 
Furthermore, hot unplug can only be done in coarse-grained units. For example, it is not possible to unplug 1.5 vCPUs. 

\vspace*{-7pt}
\subsection{Hybrid Deflation Mechanisms}
Both transparent and explicit deflation have advantages and disadvantages. 
Explicit deflation---by virtue of being visible, allows the OS and applications to gracefully handle resource deflation. 
However, deflation can only be done in coarse-grained units and has a safety threshold. 
Transparent deflation can be done in more fine-grained slices and has a much broader deflation range than explicit deflation.
It does not require any guest OS support but can impose a higher performance penalty since the OS and applications do not know that they are deflated. 

Our hybrid deflation technique combines both mechanisms to exploit the advantages of each. 
Initially, a VM is deflated using explicit deflation until its safety threshold is reached for each resource. 
From this point, transparent deflation is used for further resource reclamation to extract the maximum possible resources from the VM under high resource pressure. 
Figure~\ref{fig:hybrid-code} presents the high-level pseudo-code of our hybrid deflation approach. 
The key challenge is to determine the hot unplug safety threshold so as to switch over from explicit to  transparent deflation.

\begin{figure}[ht]
  \centering
\vspace*{-10pt}
\begin{lstlisting}[language=Python, numbers=left, frame=single, basicstyle=\scriptsize\sffamily]
def deflate_hybrid(target):
  hotplug_val = max(get_hp_threshold(), round_up(target))
  deflate_hotplug(hotplug_val)
  deflate_multiplexing(target) 
\end{lstlisting}
\vspace*{-6pt}  
  \caption{Pseudo-code for hybrid resource deflation. }
  \label{fig:hybrid-code}
\vspace*{\captionspace}
\end{figure}

For deflating CPUs, we first set the hotplug target by rounding up the target number of vCPUs (line 2 in Figure~\ref{fig:hybrid-code}).  
Then the cgroups based CPU multiplexing deflation can deflate the VM the rest of the way. 
The hotplug operation may not always succeed in removing all the CPUs requested---the guest OS unplugs the CPU only if it is safe to do so. 
If the number of reclaimed CPUs via hotplug is less than the number requested, then the multiplexing-based CPU deflation takes up the slack. 
When deflating memory, we set the hotplug threshold by using the guest OS's resident set size (RSS)---since unplugging memory beyond the RSS results in guest swapping, and we presume that it is safe to unplug as long as the VM has more memory than the current RSS value. 

\begin{figure}[t]
  \centering
  \includegraphics[width=0.3\textwidth]{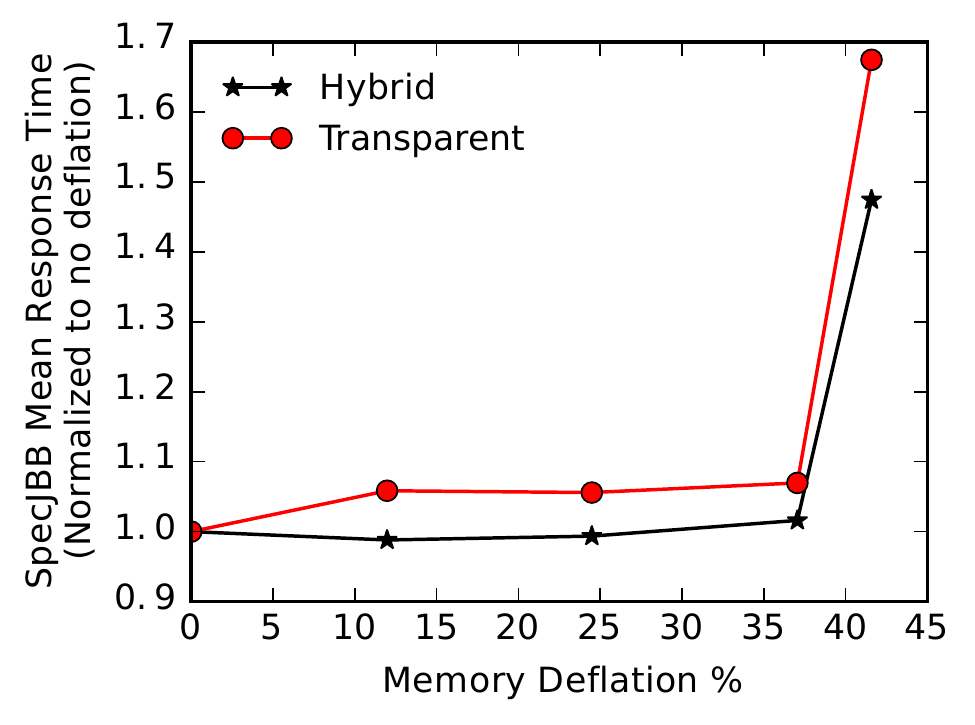}
  \vspace*{\captionspace}
  \caption{Performance of SpecJBB 2015 with transparent and hybrid memory deflation.}
  \label{fig:jbb-mem}
  \vspace*{\captionspace}
\end{figure}

Our hybrid deflation mechanisms can be used to reclaim significant amounts of CPU, memory, and I/O resources from applications. When deflating memory, hybrid deflation allows the guest OS to hot-unplug unused memory, which can improve performance, as shown in Figure~\ref{fig:jbb-mem}.
The figure shows the mean response time with the SpecJBB 2015 benchmark, and we see that the performance with both transparent and hybrid deflation is largely unaffected up to 40\% deflation, and hybrid deflation improves performance by about 10\%.
Additional results with CPU deflation and with other applications are presented later in Section~\ref{sec:eval}. 

 \vspace*{\captionspace}

\vspace{4pt}
\section{Cluster Deflation Policies}
\label{sec:policies}

In this section, we describe how the mechanisms discussed in the previous section can be
used to implement cluster-level deflation policies. We assume a cloud resource management 
framework that multiplexes physical servers in the cluster across two pools of VMs: 
non-deflatable higher-priority VMs and deflatable lower-priority VMs. When there is surplus capacity in the cluster, the cloud manager allocates these resources to lower priority VMs (without deflating them). When demand from higher-priority VM causes resource
pressure, resources from lower priority VMs are reclaimed using deflation and re-assigned to higher priority VMs.  Below, we  describe  \emph{policies} for doing so that determine how much each VM is actually deflated by, and under what conditions. 
Our policies assume the worst-case linear correlation between deflation and performance, as shown by Figures \ref{fig:util-all} and \ref{fig:underalloc}. 
Which policy to apply we leave up to cloud providers as they have different trade-offs and capabilities that we discuss in Section \ref{sec:clust-policy-eval}.
The policies we propose are implemented at the level of a physical server. 
That is, the deflation of a VM is determined by the ``local'' conditions and the resource profiles of co-located VMs.

\vspace*{-5pt}
\subsection{Server-level Deflation Policies}
\label{subsec:server-deflation}

Our system uses three policies for deflation--proportional, priority-based and deterministic---that we describe below.

\subsubsection{Proportional Deflation}
In the simplest case, we assume that all VMs that fall into two broad classes:  high-priority non-deflatable VMs (aka on-demand), and low-priority deflatable VMs.
A server may host VMs of both classes.

Proportional deflation involves deflating each low priority VM in proportion to its original maximum size. More formally, suppose we need to reclaim $R$ amount of a particular resource (CPU, memory, etc.) from $n$ deflatable VMs, and suppose $M_i$ is the original undeflated allocation of that resource allocated to VM $i$.
The proportional deflation policy reclaims $x_i$ amount from each VM $i$:
 \vspace*{-3pt}
\begin{equation}
x_i = M_i - \alpha_1 \cdot M_i, 
\label{eq:simple-prop}
\vspace*{-3pt}
\end{equation}
where $\alpha_1$ is determined by the constraint that $\sum x_i = R$, and is given by $\alpha_1 = 1-({R}/{\sum_i^n M_i})$.
Intuitively, we want VMs to deflate in proportion to their size, to avoid excessively deflating small VMs.
Note that a new incoming VM may be deflatable, and is included in the pool of $n$ deflatable VMs, and can thus start its execution in a deflated mode under high resource pressure conditions.

This simple proportional deflation policy forms the basis of more sophisticated policies for addressing various cluster management requirements.
For instance, some VMs may have a ``limit'' to their deflatability or QoS minimum requirements if deflated by more than, say, 80\%.
Applications can provide these requirements to the cluster on provisioning. The cluster manager enforces the minimum resource allocation ($m_i$) with proportional deflation policy, and reclaim resources from each VM using the following relation:
 \vspace*{-3pt}
\begin{equation}
  x_i = (M_i - m_i) - \alpha_2 \cdot ({M_i-m_i}) \\
\label{eq:min-prop}
\vspace*{-3pt}
\end{equation}

The proportional deflation is performed for each resource (CPU, memory, disk bandwidth, network bandwidth) individually.
Enforcing the minimum resource allocation limits can minimize application performance degradation, but can reduce the overcommitment (and possibly revenue) of cloud platforms.

\vspace*{-2pt}
\subsubsection{Priority-based Deflation}
Since the impact of deflation is application dependent, a cloud platform can offer multiple classes of deflatable VMs.
These priority levels influence how much each VM is deflated by, and can be offered by cloud providers at different prices. 
These priority classes can be chosen by the user based on their price sensitivity and application characteristics. 

The proportional deflation policy (Equation~\ref{eq:simple-prop}) can be extended to incorporate priorities through a weighted proportional deflation framework.
Let $\pi_i\in (0,1)$ be the priority level of VM-$i$.
Then, 
 \vspace*{-3pt}
\begin{equation}
x_i = M_i - \alpha_3 \cdot \pi_i \cdot M_i , 
\label{eq:prio-simple}
\vspace*{-3pt}
\end{equation}
where low $\pi_i$ values indicate lower priority and higher deflatability. 

VM priorities can also be applied to determine the minimum resource allocation levels ($m_i$) of the VMs.
Intuitively, VMs with a higher priority ($\pi_i$) have a lower deflation tolerance, and thus larger $m_i$ values.
For instance, cloud platforms can determine the VM's minimum resource allocation level as: $m_i = \pi_i \cdot M_i$, and we can then extend the  minimum-level-aware deflation (Equation~\ref{eq:min-prop}) with weighted proportional deflation:

\vspace*{-8pt}
\begin{equation}
  x_i = (M_i - \pi_i M_i) - \alpha_4 \cdot \pi_i ({M_i-\pi_i M_i})
\label{eq:prio-all}
\vspace*{-3pt}
\end{equation}

\subsubsection{Deterministic Deflation}
With the above proportional deflation policies, a VM's deflation level is determined dynamically based on the local resource pressure on the server. 
In some cases, cloud platforms and applications may require a more deterministic deflation policy, that only deflates VMs to a pre-specified level. 
VM priorities can be used for determining the deflation levels of VMs---with higher priorities ($\pi_i$) indicating lower deflation.
In this case, deflation is binary: either the deflatable VMs are allocated 100\% of their resource allocation ($M_i$), or $\pi_i\cdot M_i$.
In case of multiple deflatable VMs on a server, VMs are deflated in decreasing order of $\pi_i$'s until sufficient resources are reclaimed to run the new VM. 

\noindent \textbf{Reinflation:} Both our proportional and priority-based  policies can also reinflate previously deflated VMs when additional resources become available. 
When $R_{\text{free}}$ additional resources have become available, we reinflate VMs proportionally by setting $R = -R_{\text{free}}$ in equations~\ref{eq:simple-prop},~\ref{eq:min-prop},~\ref{eq:prio-simple},~\ref{eq:prio-all}, and effectively run the proportional deflation backwards in all the cases. 
For deterministic deflation, the highest priority VMs are reinflated first.

\vspace*{-7pt}
\subsection{Deflation-aware VM Placement}
\label{subsec:vm-placement}
\vspace*{-4pt}

The initial placement of VMs onto physical servers also affects their deflation.
Conventionally, for non-deflatable VMs, bin-packing based techniques are used by cluster managers to place VMs onto the ``right'' server in order to minimize fragmentation and total number of servers required. 
This is often solved through multi-dimensional bin-packing lens.
The VM's CPU, memory, disk and network resource needs as well as the resources available on each server are multi-dimensional vectors.  Policies such as best-fit or first-fit can be used to choose a specific server.  
We  use the notion of ``fitness'' for placing  VMs onto a server. 
Similar to~\cite{tetris}, we use the {\em cosine similarity} between the demand vector and the availability vector to determine fitness:
$  \text{fitness}(\mathbf{D}, \mathbf{A_j}) = \frac{\mathbf{A_j} \cdot \mathbf{D}}{|\mathbf{A_j}||\mathbf{D}|}$.
Here, $\mathbf{D}$ is the demand vector of the new VM, and $\mathbf{A_j}$ is the resource availability vector  of server $j$. If $A_j = 0$, i.e. there are no available resources, a small value $\epsilon$ can be added to it, or the server can be removed from consideration, to prevent division by 0.
The availability vector is given by $A_j = \text{Total}_j - \text{Used}_j + (\text{deflatable}_j / \text{overcommitted}_j)$, where $\text{deflatable}_j$ is the maximum amount of resources that can be reclaimed by deflation and $\text{overcommitted}_j$ is the extent of the deflation already done. 
By evaluating all severs and considering their level of overcommitment, this approach prefers servers with lower overcommitment, and thus achieves better load balancing. 

\vspace*{-7pt}
\subsubsection{Placement With Cluster Partitions.}
The above VM placement approach results in VMs of different priority levels sharing physical servers.
This ``mixing'' can be beneficial and improve overall cluster utilization, since lower priority VMs can be deflated to make room for higher priority VMs. 
However, increasing the number of co-located deflated VMs can potentially result in higher performance interference (aka noisy neighbor effect).

While performance interference can be mitigated through stronger hypervisor and hardware-level isolation techniques, it can also be addressed by VM placement.
The key idea is to \emph{partition} the cluster into multiple priority pools, and only place VMs in their respective priority pools.
Within a pool, we use the bin-packing approach for deflatable VMs and continue to use either proportional or deterministic deflation policies on the individual servers.
The size of the different pools can be based on the typical workload mix.

Thus, higher priority VMs will generally run on servers with lower overcommitment and lower risk of performance interference, and lower priority VMs face higher risk of overcommitment.
This approach also allows cloud operators to limit and control the distribution of overcommittment of different servers, which reduces the risk of severe performance degradation due to overcommitment.

A possible downside of cluster partitions is that if a partition becomes ``full'' even after deflating all its VMs to their maximum limits, new VMs may have to be rejected using the admission control mechanism.
This can reduce cluster overcommitment and revenue. 

\vspace*{-10pt}
\subsubsection{Pricing Considerations}

Our work assumes that deflatable VMs are priced differently from traditional on-demand VMs. Similar to preemptible VMs, a cloud provider may choose to offer deflatable VMs 
at fixed discounted prices (e.g., at 60-80\% discount). The cloud provider may also price
deflatable VMs based on priority levels, where the priority level determines the proportion by which VM can be deflated and also the  discount in the price. Finally, the cloud provider may use variable pricing where the deflatable VM is billed based on the actual allocation of resources over time, with lower prices charged during periods of deflation. The different pricing policies, when combined with placement and server-level deflation policies, result in different levels of application performance, cluster utilization, and revenue. These tradeoffs are presented in the evaluation section.

\vspace*{-6pt}
\section{Implementation}

\label{sec:impl}

We have implemented all the deflation mechanisms and policies discussed in Sections \ref{sec:mechanisms}-\ref{sec:policies} as well as deflation-aware web applications, as part of a deflation-aware cluster manager framework. 
Our system is comprised of two main components (see Figure \ref{fig:defl-over}).
A centralized cluster manager implements and invokes the VM placement policies and generally controls the global-state of the system. 
In addition, we run local deflation controllers that run on each server. 
These local controllers control the deflation of VMs by responding to resource pressure, by implementing the proportional deflation policies described in~\cref{sec:policies}. 
Both the centralized cluster manager and the local-controllers are implemented in about 4,000 lines of Python and communicate with each other via a REST API. 

\noindent {\bf Deflation Mechanisms.}
Our prototype is based on the KVM hypervisor~\cite{kivity2007kvm}, and uses the libvirt API for running VMs and for dynamic resource allocation required for deflation. 
Our hybrid resource deflation mechanisms presented in~\cref{sec:mechanisms} are implemented by the per-server local controller. 
CPU and memory hot-plugging (and unplugging) are performed via QEMU's agent-based hotplug.
Hotplug commands are first passed to the user-space QEMU agent, which then forwards them to the guest OS kernel. 
Thus, the guest OS is made aware of the deflation attempt, and knows the unplug is not due to hardware-failure, and  allows the hotplug to be ``virtualization friendly''. 
For example, if the guest kernel cannot safely unplug the requested amount of memory, the hot unplug operation is allowed to return unfinished. 
In this case, the memory reclaimed through hot plug will be lower, but the safety of the operation is maintained. 

For hypervisor level multiplexing of resources, we run KVM VMs inside cgroups containers, which allows us to multiplex resources. 
For CPU multiplexing, we adjust the cgroups cpu shares of the VM through libvirt's cgroups API. 
For transparent memory deflation, we adjust the VM's physical memory usage by setting the memory usage of the cgroup (\textsf{mem.limit\_in\_bytes}).  
Disk and network bandwidth are also dynamically adjusted via  libvirt API's.

\noindent \textbf{Deflation Policies.} The server-level deflation policies are implemented by a local deflation controller on each server, which maintains and manager all aspects of the  server's resource allocation state, and determines deflation amounts of different VMs.
Each server updates the central master about all changes in server utilization after every deflation event. 
New VMs are placed on servers using a three-step approach.
First, the centralized cluster manager finds the ``best'' server for the VM based on the VM size and utilizations of all servers.
The second step involves the server computing the deflation required to accommodate the new VM.
If this violates any resource constraint, then the server rejects the VM.
Finally, the actual deflation is performed and the VM is launched.

\noindent \textbf{Deflation-aware Web Cluster:}
When running web clusters on deflatable VMs, the load balancer can be made deflation aware for improved performance. The load balancer can adjust the number of requests sent to a VM based on its deflation level.
We implement a deflation-aware load-balancing policy in HAProxy~\cite{haproxy}. 
We have modified HAProxy's Weighted Round Robin algorithm by dynamically changing the weights assigned to the different servers based on the current deflation level, which adjusts the number of requests sent to each server based on the ``true'' resource availability. 
The load balancer changes are implemented in Python and Kotlin in a total of 300 LOCs and are wrapped in a Docker container.

\vspace*{-6pt}
\section{Experimental Evaluation}
\label{sec:eval}

In this section, using  testbed experiments and simulation, we show the performance of deflatable VMs, and focus on answering the following questions:
\begin{enumerate}
\item What is the performance of interactive applications when deployed on deflatable VMs? 
\item What is the impact of deflation policies on cluster utilization, application throughput, and cloud revenue?
\end{enumerate}

\vspace*{-8pt}
\subsection{Evaluation Environment}
\vspace*{-5pt}
\subsubsection{Web-based interactive applications.}

\begin{figure}[t]
  \centering
  \includegraphics[width=0.45\textwidth]{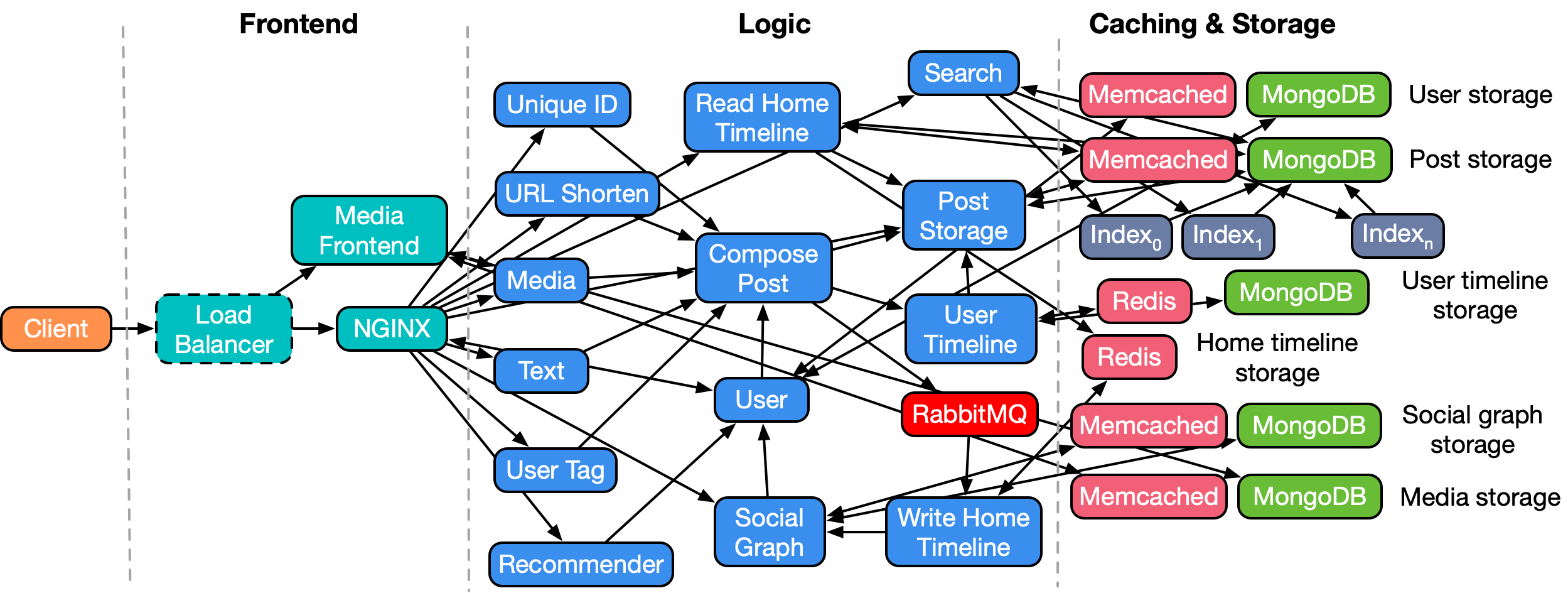}
  \vspace*{\captionspace}
  \caption{The micro-service architecture of the social network application used in our evaluation (Courtesy of~\cite{gan2019open}).}
    \label{fig:death}
  \vspace*{\captionspace}
  \vspace*{-0.2cm}
\end{figure}

We use two interactive applications to evaluate deflation on real-world web workloads:

\noindent  \textbf{Wikipedia:} We replicate the German Wikipedia on our local testbed. 
We choose the German Wikipedia as it is the second most popular Wikipedia in terms of number of views---with more than 720000 page views per hour---, and the fourth most in terms of number of articles---with more 2.25 Million articles~\cite{WikiStats}.
We setup a KVM VM with MediaWiki, MySQL database, Apache HTTP webserver, and Memcached. 
Our workload generator randomly selects from the top 500 largest pages (page sizes ranging from 0.5--2.2 MB). 

\noindent \textbf{DeathStarBench} is a recently released benchmark that implements different applications using the microservice architecture~\cite{gan2019open}.  We evaluate the benchmark's  social networking application, which consists of 30 microservices (Figure~\ref{fig:death}) built using Redis, Memcached, MongoDB, RabbitMQ, Nginx, Jaeger, and other custom made services that provide the required functionality. 
We run each micro-service runs in a separate Docker container using using Docker swarm.
We use a workload generator based on wrk2~\footnote{https://github.com/giltene/wrk2} for evaluating the overall application performance.

\begin{figure*}
\begin{minipage}[c]{0.3\linewidth}
\includegraphics[width=\linewidth]{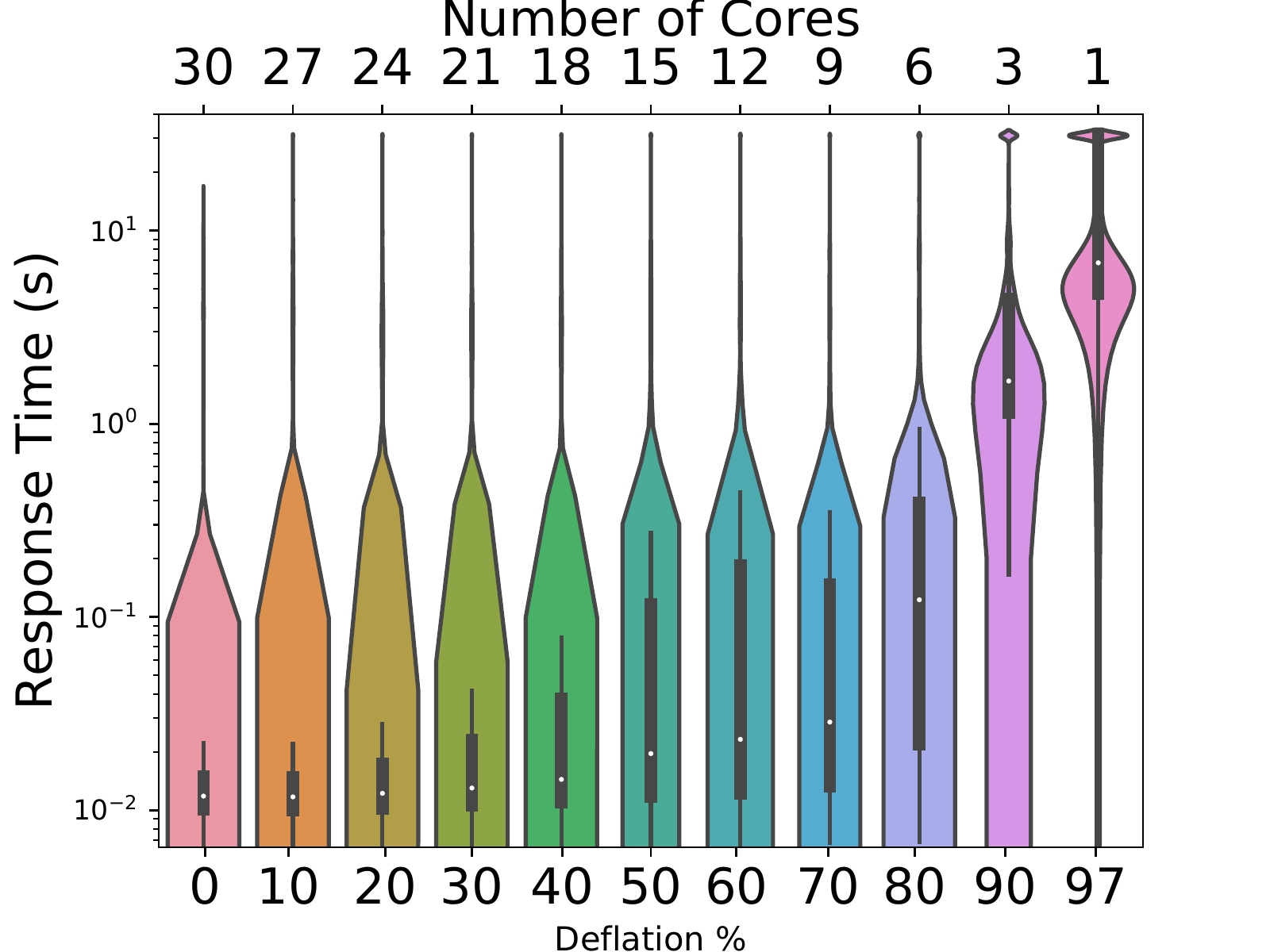}
\vspace{-0.3in}
  \caption{Wikipedia response times with CPU deflation.}
  \label{fig:wikiviolin}
 \end{minipage}
\hfill
\begin{minipage}[c]{0.3\linewidth}
\includegraphics[width=\linewidth]{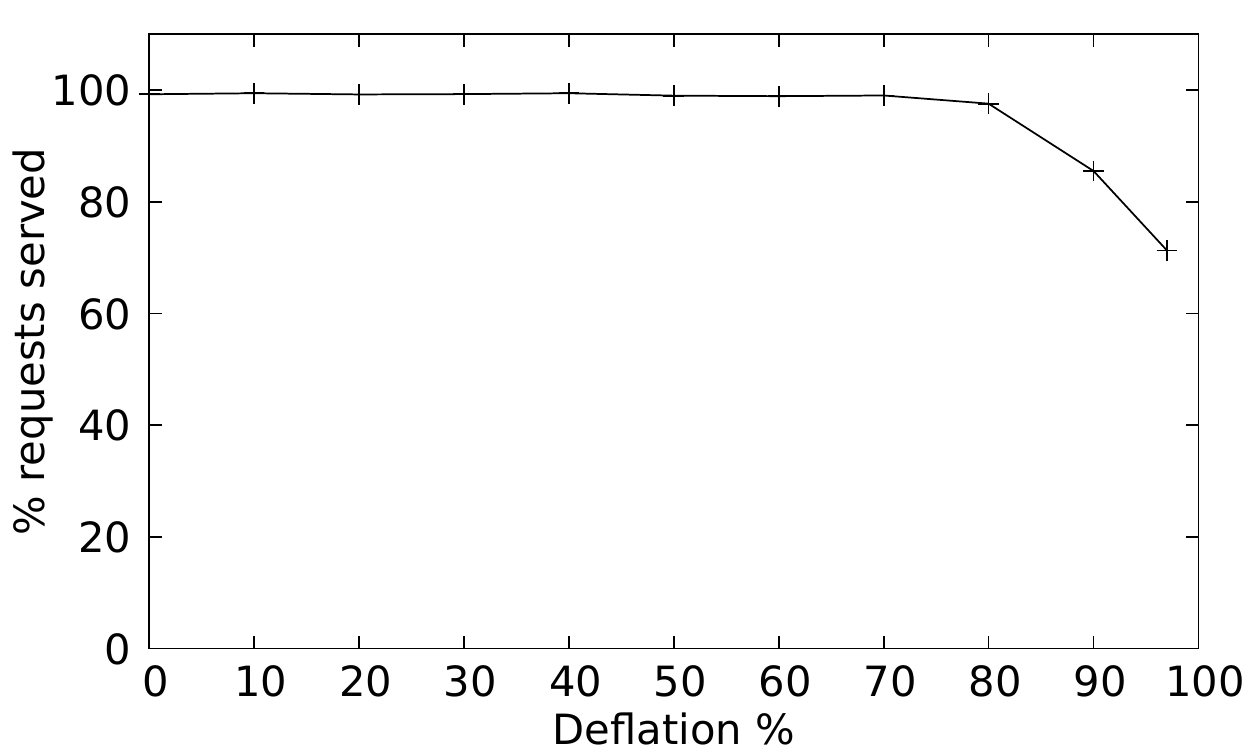}
 \vspace{-0.2in}
 \caption{Almost all Wikipedia requests are served till 70\% deflation.}
   \label{fig:wikiloss}
  \end{minipage}\hfill
\begin{minipage}[c]{0.3\linewidth}
\includegraphics[width=\linewidth]{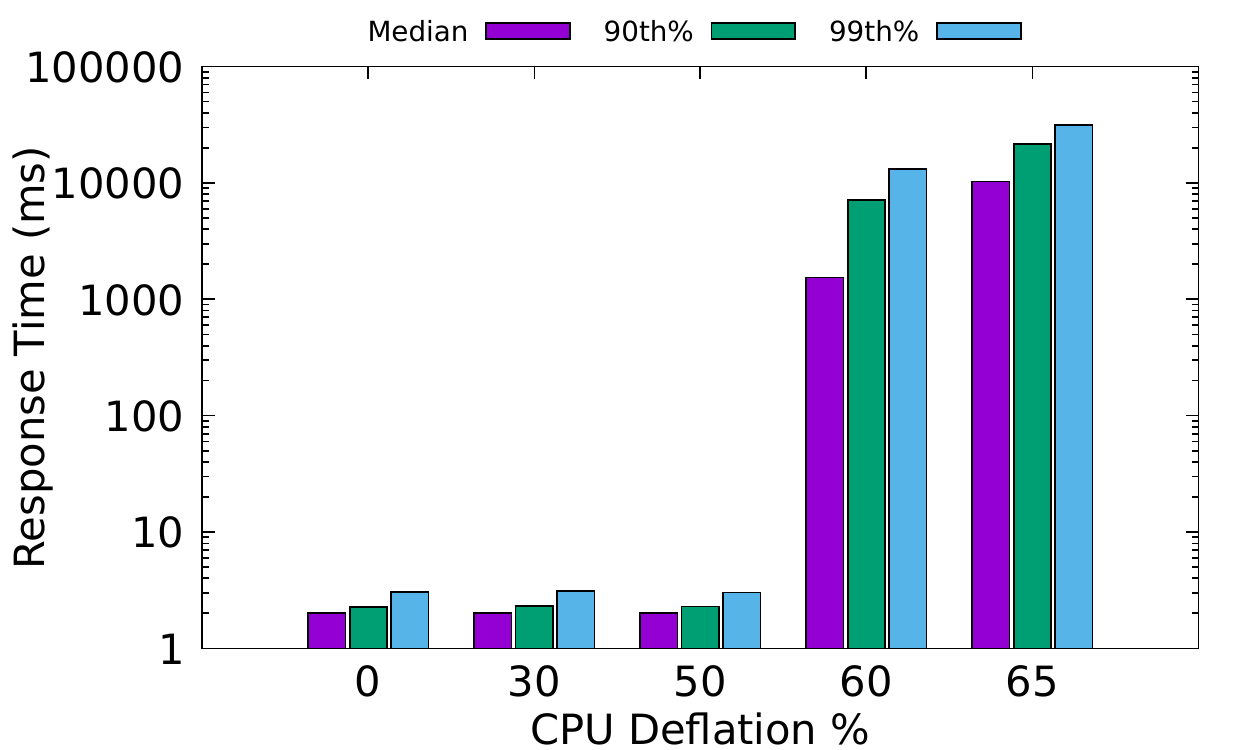}
\vspace{-0.2in}
  \caption{Response times for the social media application.}
 \label{fig:uservices}
\end{minipage}\vspace*{-18pt}
\end{figure*}

\vspace*{\captionspace}
\vspace*{4pt}
\subsubsection{Cluster-level simulation framework.}

To analyze various cluster-level deflation policies, we have developed a trace-driven discrete event simulation framework that allows us to understand the impact on application and cloud-level metrics. 
The simulation framework is written in Python in about 2000 lines of code, and implements our VM deflation and pricing policies, and allows large-scale simulations with different policy and workload combinations. 
We use the Azure VM-level dataset to determine the starting and stopping times of VMs, their size (aka resource vectors), and CPU utilization history. 
We also use the VM metadata such as VM category (batch, interactive, unknown), and the 95-th percentile CPU utilization to determine priority levels for our priority-based deflation policies. 
The simulation framework allows us to determine the deflation levels of VMs, preemptions in the cluster, and also correlate VM's dynamic resource allocation with its CPU utilization time-series to determine the performance impact of deflation.
For the simulation-based cluster-level experiments, we primarily focus on the effect of deflation on the cloud provider.
This complements our application-focused performance evaluation done using web services in the next subsection, as well as prior work on deflation~\cite{deflation-eurosys19} that looked at performance of distributed applications under deflation.

Given our focus on deflatability of interactive applications, we assume that the interactive VMs in the trace are  deflatable, while the unknown and batch VMs are non-deflatable (``on-demand''). 
This translates to roughly 50\% of the VMs being deflatable.
We consider each VM's CPU core count and memory size for bin-packing as well as all deflation policies. 
We determine VM priorities based on their 95-th percentile CPU usage and use 4 priority levels.
We show results on a randomly sampled trace of 10,000 VMs, which require a cluster of 40 servers each with 48 CPUs and 128 GB RAM.
For simulating varying degrees of cluster overcommitment, we first find the minimum cluster size capable of running all VMs without any preemptions or admission-controlled rejections. 
We then vary and increase the overcommitment by reducing the number of servers and use the same VM-trace throughout for all the experiments. 
We do not look at the impact on individual application performance in a cluster settings for two reasons 1) the cluster level impact of deflation was examined in \cite{deflation-eurosys19} and 2) we want to focus on the effect of our deflation policies on large-scale cluster management.

\vspace*{\captionspace}
\vspace*{-7pt}
\subsection{VM deflation of Web services}
\vspace*{\captionspace}
\vspace*{7pt}
Our first set of experiments aim to measure the effect of \emph{transparent} deflation on the performance of different types of web services, and how the reduction in resource allocation can be mitigated by well-engineered web applications.

\noindent \textbf{Multi-tiered Applications.}  In order to evaluate the effects of deflation on the QoS  of multi-tiered services, we use the German Wikipedia replica running on a VM with 30 CPU cores, and 16 GB of memory.
We  subject it to a mean load of 800 requests/s selected randomly from the 500 largest pages.
We set the request time out period to 15 seconds, and consider that requests that take longer are dropped, or no longer interesting to the users.
We progressively deflate the VM's CPU for this CPU-bound application. 
Figure~\ref{fig:wikiviolin} shows a violin plot of the distribution of the response times of the requests at each deflation level, with the y-axis in log-scale.
As shown, the response time does not increase significantly until the deflation increases above 70\%---even though the average CPU usage at 50\% deflation is 100\%.
We find that the average response time for the application with no deflation is 0.3s, with 50\% deflation is 0.45s, and with 80\% deflation is 0.6s---which is $2\times$ the undeflated response time. 
The 99th percentile response time is 6.8s for no deflation, and increases by only 43\% to 9.74s even at 80\% deflation.
We also find that, even when deflated to a single core, the application did not crash when serving a load of 800 req/s.
This leads us to believe that many well architected web services tolerate deflation well, with a disproportionately small performance penalty.  This observation is further reinforced by
Figure~\ref{fig:wikiloss}, which shows the percentage of requests served with different deflation settings.
Similar to our previous result, we see that noticeable request loss rates occur only after 70\% deflation.

\noindent \textbf{Micro-service based Applications.} We next evaluate the impact of transparent deflation on micro-service based applications. Figure~\ref{fig:death} shows the architecture of the social networking application described previously.
The application microservices can be classified based on their functionality into three logical classes that are similar to multi-tiered applications, namely, frontend microservices, logic microservices, and finally, caching and storage microservices. In the social networking service used, there are three frontend microservices, 15 logic microservices, and 12 backend microservices.
In our deflation experiment, we deflate all microservices except for the databases, i.e., we deflate all frontend and logic microservices, and the four memcached microservices from the backend, deflating a total of 22 microservices out of 30.
We start by allocating a maximum limit of 2 cores per microservice, and a minimum of 0.05 CPUs for each container.
Each container is allocated 800MB of memory.
We use the workload generator to generate 500 requests per second, and deflate the 22 microservices by 30\%, 50\%, 60\% and 65\%.
Figure~\ref{fig:uservices} shows the  median, 90th\%, and 99th\% response times in milliseconds.
We again see that the service can be deflated by up to 50\% with no performance losses.
Beyond this level, the degradation in QoS and response time  is more
abrupt than the the multi-tiered Wikipedia case, likely due to the higher
communication- and coordination-intensive nature of the application. 

\vspace*{-8pt}
\subsection{Deflation-aware Web Load Balancing}
\vspace*{-4pt}
Next, we evaluate the effect of \emph{explicit} deflation on clustered web services.
To do so, we compare the performance of using vanilla HAProxy~\cite{haproxy} with our modified deflation-aware HAProxy. 
We run three replicas of the German Wikipedia application behind HAProxy.
Each instance starts with 10 vCPU cores, and 10 GB of memory.
We assume that two of these instances are running on deflatable VMs , and the third runs on a non-deflatable VM.

\begin{figure}[t]
  \centering
  \includegraphics[width=0.38\textwidth]{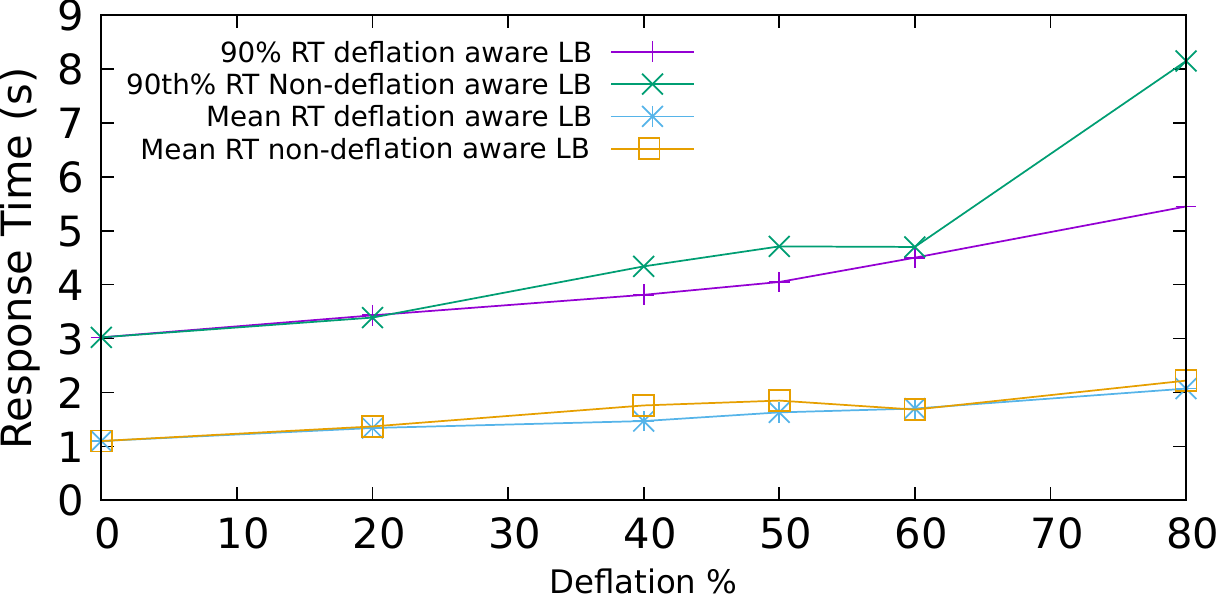}
  \vspace*{\captionspace}
  \caption{Our deflation-aware load balancer yields lower response times even at high deflation levels.}
  \label{fig:lbc}
  \vspace*{\captionspace}
\end{figure}

We generate an average load of 200 requests/s and deflate the two deflatable VMs equally.
Our deflation-aware load balancer attempts to masks the impact of deflation by changing the server weights based on the deflated number of vCPUs, causing more requests to be sent to the third undeflated replica.
Figure~\ref{fig:lbc} shows the average and 90th percentile response times for the unmodified and deflation-aware load balancers. We see that the deflation-aware load balancer yields 15 to 40\% lower tail latency at high deflation levels of 40 to 80\% when compared 
to vanilla load balancing; mean response times are also lower or comparable as shown in the figure.

\vspace*{-9pt}
\subsection{Impact Of Cluster Deflation Policies}
\vspace*{-4pt}
\label{sec:clust-policy-eval}
We now evaluate the effect of VM deflation at a cluster level using trace-driven simulations.
We are interested in the differences with current transient server offerings that rely on preemptions, and the impact of the different deflation policies on cluster overcommitment, VM performance, and cloud revenue.

\begin{figure}[t]
  \centering
  \includegraphics[width=0.35\textwidth]{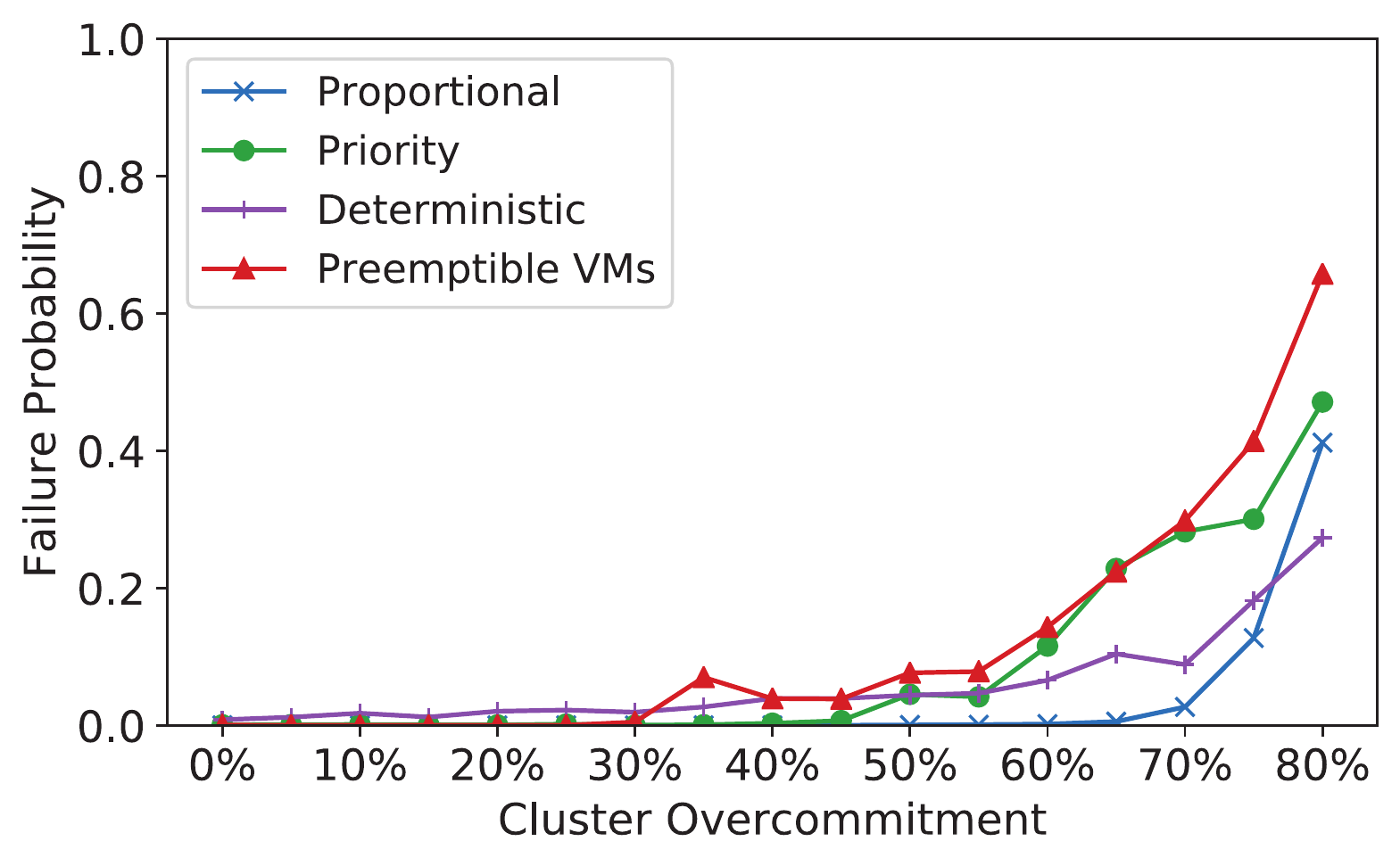}
  \vspace*{\captionspace}
  \caption{Failure probability with deflation remains very low even for high cluster overcommitment.}
  \label{fig:preemptions}
  \vspace*{\captionspace}
\end{figure}

\vspace*{-4pt}
\subsubsection{Eliminating Preemptions.}
VM deflation is intended to eliminate preemptions, which are detrimental to interactive applications because they cause downtimes.
Currently, cloud operators preempt low-priority VMs when there is high resource pressure, which increases at high cluster overcommitment levels.
Figure~\ref{fig:preemptions} shows the failure probability for low-priority VMs under different overcommitment levels. Failure probability represents the probability of failure to reclaim sufficient resources from deflatable VMs due to "too much" overcommitment; for
traditional preemptible instances, it is same as preemption probability. 
Even at 70\% overcommitment, the failure probability is below 1\% for proportional deflation, compared to 35\% for preemptible VMs. From a provider standpoint, this implies that they can reclaim the desired amount of resources via deflation with $>$0.99 probability. The priority-based and deterministic  deflation policies have higher failure probability than proportional but  still below preemptible VMs. More broadly, this result shows
that a judicious choice of overcommitment level (of as much as 50\%) allows the provider
to eliminate preemptions and use deflation to reclaim the necessary resources under resource pressure.

\noindent \textbf{Result:} \emph{Deflatable VMs have very low  probability of resource reclamation failure even when the overcommitment is as high as 50\%}

\vspace*{-5pt}
\subsubsection{Throughput.}

\begin{figure}[t]
  \centering
    \includegraphics[width=0.35\textwidth]{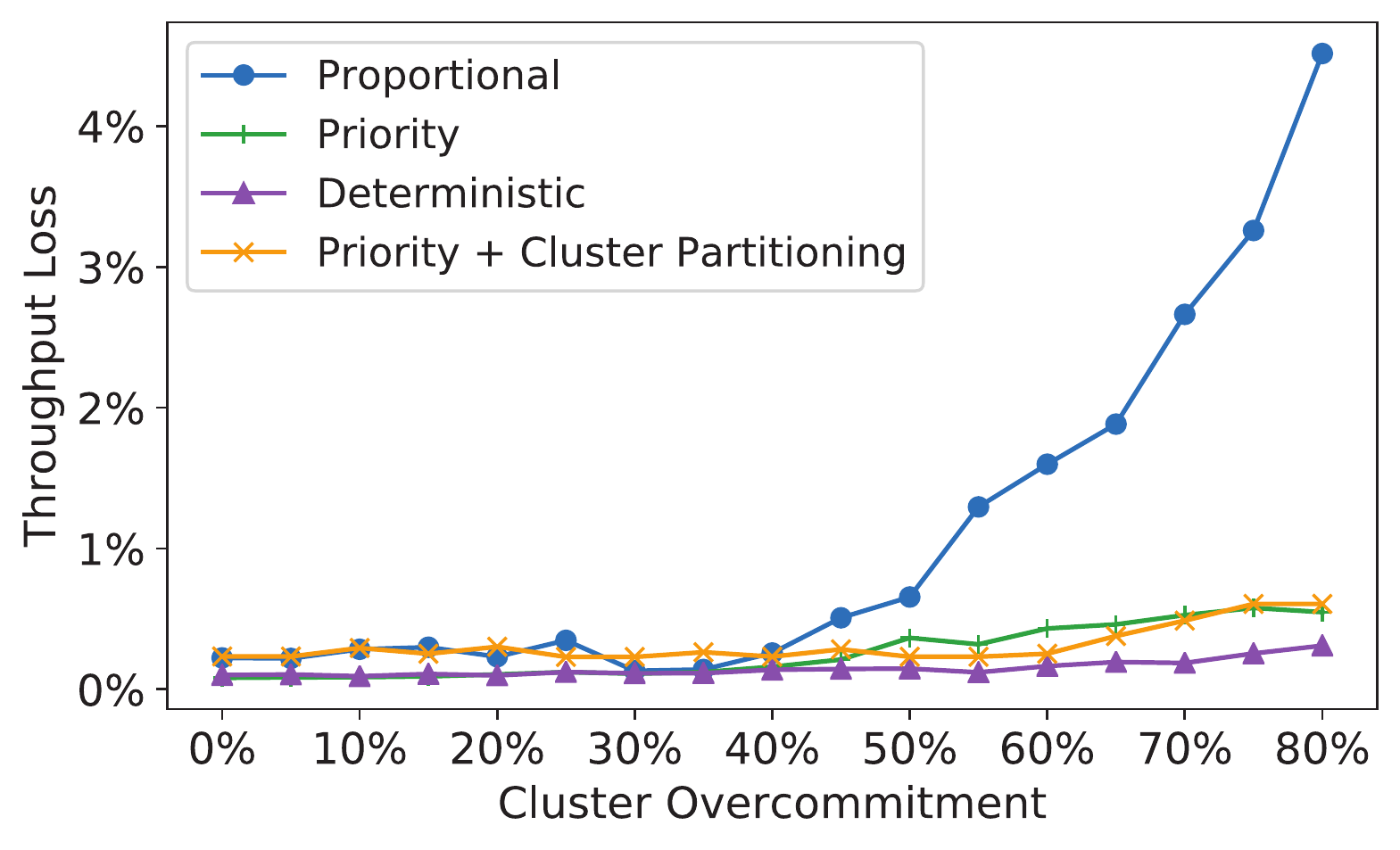}
    \vspace*{-15pt}
    \caption{Decrease in throughput of deflatable VMs is low even at high overcommitment.}
    \label{fig:tputloss}
\vspace*{\captionspace}
\vspace*{-5pt}
\end{figure}

While deflation can eliminate preemptions, it comes with an important tradeoff: 
the reduction in resource allocation due to overcommitment can reduce application performance and throughput. 
We examine the effect of deflation on VM performance at a cluster level, using the CPU-traces of the Azure VMs. 
Note that a VM's deflation is \emph{dynamic} and based on the time-varying resource pressure conditions as VMs are launched and terminated.
At a given point in time, the performance depends on the deflation and the VM's resource utilization. 
Thus if the VM is deflated when its resource (CPU) utilization is low, then we are reclaiming unused resources (i.e., slack), and there should be no drop in throughput.
The loss in throughput only occurs when a VM is deflated below its CPU usage, and is proportional to the total underutilization (area under the curve of Figure~\ref{fig:underalloc} in Section~\ref{sec:feasibility}. 
Based on this principle, Figure~\ref{fig:tputloss} shows the decrease in throughput for the different deflation policies at varying  overcommitment levels. 

We see negligible reduction in throughput below 40\% overcommitment, and a 1\% reduction at 50\% overcommitment. 
Even at 80\% overcommitment, the loss in throughput is below 5\% for all deflation policies.
We note that this is fundamentally due to the low utilization of VMs of the Azure VMs (especially interactive VMs), as was shown earlier in Figure~\ref{fig:bp-over-thresh}. 
Additionally, the average VM deflation is \emph{not} equal to the cluster overcommitment but is significantly lower. 
Our cluster was provisioned for the \emph{peak} load, and furthermore, deflatable VMs significantly improve the bin-packing efficiency by allowing the cluster manager to slightly adjust VM allocations to make room for new VMs that would have otherwise not fit and required an additional server.

The priority-based and deterministic deflation policies take into account the VM's anticipated utilization levels by using their 95 percentile CPU usage to determine the deflation priority and the minimum allocation levels.
Thus, high utilization VMs are deflated less, which reduces their loss in throughput compared to simple proportional deflation.
Thus, we see that adding priorities can reduce the loss in throughput by an order of magnitude. 
When we place VMs into dedicated cluster partitions based on their priority (as described in Section~\ref{subsec:vm-placement}), Figure~\ref{fig:tputloss} also shows that incorporating partitioning does not significantly impact throughput loss. 
Cluster-partitioning is thus a viable technique that can be used by cloud operators to minimize the risk of performance interference among deflatable VMs of different priorities.

Interestingly, deterministic deflation, which deflates VMs in their priority order, has the lowest decrease in throughput.
This is because the proportional deflation policies (both the simple and priority-based proportional) result in deflation of \emph{all} VMs, even though the magnitude of deflation of each VM is small.
Thus, even high priority deflatable VMs are deflated, and their throughput will decrease if their CPU utilization is higher than the deflated allocation.
With deterministic deflation, the lower priority VMs (with lower 95 percentile CPU usage) are penalized more, but the average cluster-wide throughput loss is reduced.

\noindent \textbf{Result:} \emph{Deflatable VMs allow clusters to be overcommited by 80\%, and keep the performance degradation to less than 5\%.}

\noindent \textbf{Impact on Quality of Experience.}
The low average loss in throughput represents a low risk of QoS violations, since performance is affected only when the application's peak usage coincides with deflation. 
However, end-users of interactive applications may observe a perceivable drop in their quality of experience due to the jitter and the longer response times during deflation. 
Ultimately, evaluating the user experience with deflation requires user studies similar to~\cite{skype-study}, and is a potential candidate for future work. 
End-users can be alerted with a ``degraded mode'' warning during periods of high deflation, similar to downtime indicators for popular web services. 
Finally, we note that distributed applications can also run on a mix of non-deflatable and deflatable VMs with different priorities (similar to~\cite{spotweb-hpdc19}), and reduce the risk of QoS violations even further.

\subsubsection{Cloud Revenue.}

\begin{figure}[t]
    \centering
    \includegraphics[width=0.35\textwidth]{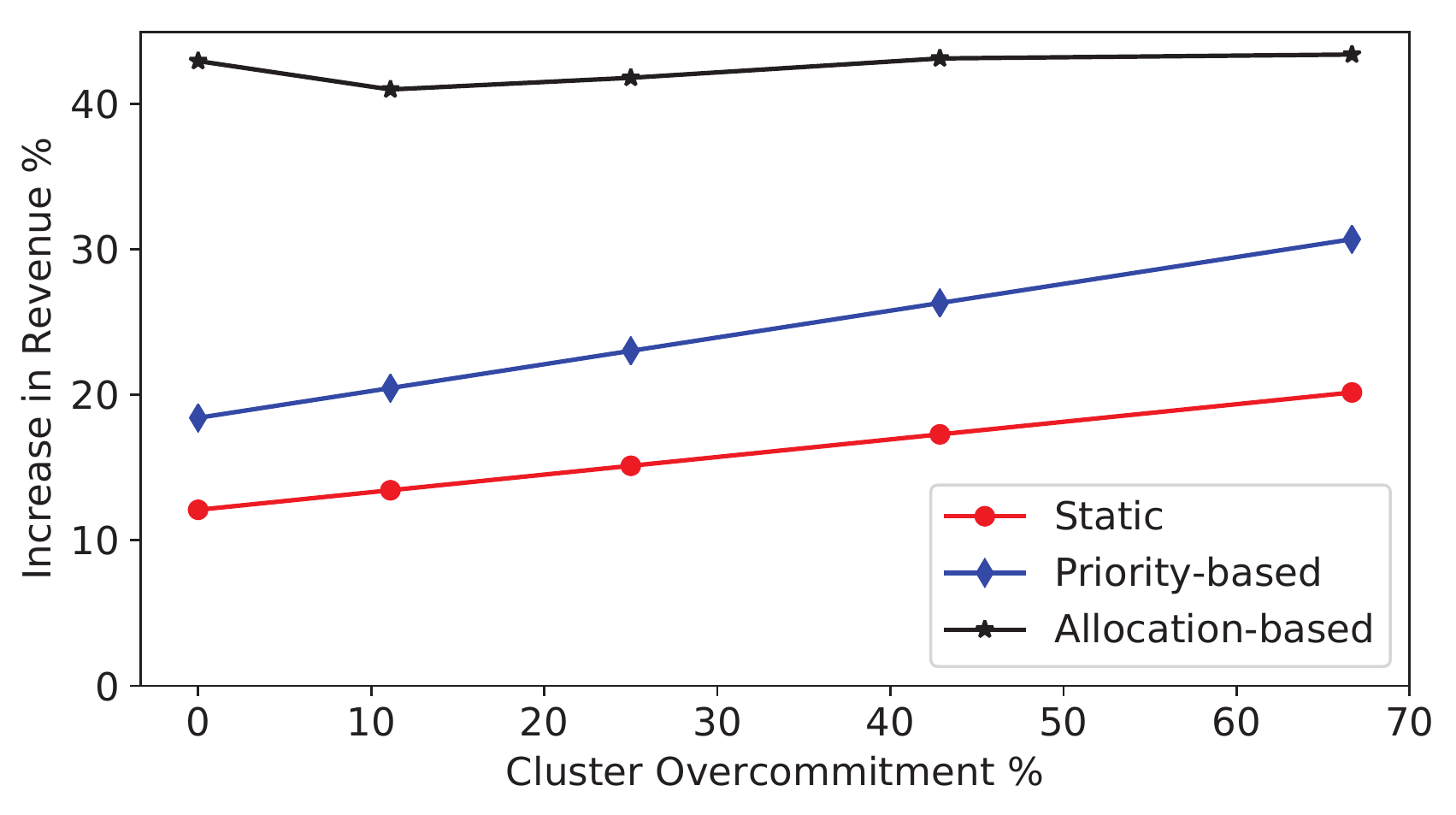}
    \vspace*{\captionspace}
    \caption{Increase in cloud revenue due to deflatable VMs.}
    \label{fig:revenue}
    \vspace*{\captionspace}
    \vspace*{-10pt}
\end{figure}

\vspace*{\captionspace}
\vspace*{3pt}
We have seen how deflatable VMs can minimize preemptions and have negligible impact on performance of interactive applications.
Since deflation allows for increased overcommitment, it provides cloud platforms the opportunity to increase their revenue on  low-priority resources.
Figure~\ref{fig:revenue} shows the increase in revenue from the low-priority (i.e., deflatable) resources, at different cluster overcommitment levels for different combinations of deflation and pricing policies. 
For ease of exposition, we assume that the static price of deflatable VMs is $0.2\times$ the on-demand price---corresponding to the discounts offered by current transient cloud servers such as EC2 spot instances, Google Preemptible VMs, and Azure Low-priority Batch VMs. 
For VMs with different deflation priorities, we set their price equal to the priority---i.e, priority-level 0.5 has price $0.5\times$ the on-demand price, etc.
We also evaluate variable allocation-based pricing which considers the actual resource allocation over time, and again price resources linearly (i.e, VMs pay half price when at 50\% allocation).

Figure~\ref{fig:revenue} shows that as the cluster overcommitment increases, the revenue with static-pricing VMs increases, and the cloud platform can increase revenue by 15\% at 60\% overcommitment. 
Having priority-based differentiated pricing significantly increases the revenue, since higher priority VMs pay more.
The priority-based pricing (when used with priority-based deflation) increases the revenue per server by $2\times$ compared to simple static pricing. 

Interestingly, the revenue with allocation-based pricing scheme, which charges VMs what they were actually allocated, does not increase with increasing overcommitment.
This is because at low overcommitment levels, VMs are not deflated and thus pay ``full price'', and as the overcommitment increases, there are more VMs running per server, but they are highly deflated, and thus the total revenue remains the same.

\noindent \textbf{Policy Comparison:} \emph{Deflation policies have different tradeoffs. Proportional deflation minimizes resource reclamation failure, but provides lower revenues. Priority-based deflation and pricing increases revenue, but also increases failure probability.}

\vspace*{-7pt}
\section{Related Work}

\label{sec:related}

VM deflation draws upon many related techniques and systems.

\noindent \textbf{Systems for handling transient server revocation} use a combination of fault tolerance and resource allocation  to mitigate the performance and cost effects of preemptions. 
Prior work has focused on system~\cite{spotcheck, spoton} and application~\cite{flint, exosphere, marathe2014exploiting, pado-eur17, proteus-eur17,  conductor} support for handling preemptions. 
We believe that deflatable VMs minimize the need for such middleware, and can avoid the performance, development, and deployment costs associated with preemption.

\noindent \textbf{Resource overcommitment mechanisms} have been extensively studied and optimized to allow for more efficient virtualized clusters. 
Memory overcommitment typically relies on a combination of hypervisor and guest OS mechanisms, and has received significant attention~\cite{waldspurger2002memory, amit2014vswapper, singleton}. 
Memory ballooning is another memory overcommitment technique with generally inferior performance to hotplug~\cite{fraser-ballooning-hotplug, liu2015hotplug}. 
Hotplug can also be used for reducing energy consumption~\cite{zhang2014dimmer}, since unused but powered-on RAM draws a significant amount of energy. 
CPU hotplugging can also be used to mitigate lock-holder preemption problems in overcommitted vCPUs~\cite{ding2014gleaner, ouyang2013preemptable}.
Burstable VMs~\cite{ec2-burstable, bhuvan-burstable} also offer dynamic resource allocation, but are the ``inverse'' of deflatable VMs.
The resource allocation is high by default for deflatable VMs and only reduced during resource pressure, whereas burstable VMs have low allocation by default and only ocassionally can be ``inflated'' to higher allocations.
Furthermore, burstable VMs have been restricted to CPU and I/O bursting, whereas deflatable VMs also adjust memory.

\noindent \textbf{Resource consolidation} using dynamic resource allocation~\cite{borg} and VM migration~\cite{wood2009sandpiper} is common to increase cluster utilization. 
VMWare's distributed resource scheduler~\cite{vmware-drs} uses per-VM reservations (minimum limits) and shares for dynamically allocating resources---similar to our resource-pressure based local deflation policies.  
Many approaches for performance-sensitive resource allocation among co-located VMs have been suggested~\cite{liu2014reciprocal, moldable-vms, zhou2010vmctune, stopgap-elastic, elasticity-driver-vee15}, but they assume some application performance model, which our work does not. 
VM memory allocations can be set using working-set estimation~\cite{zhang2016iballoon, chiang2013working, zhao2009dynamic}, utility-maximizing~\cite{hines2011-ginko}, or market-based approaches~\cite{agmon2014ginseng, nom-vee}.  As noted earlier, deflation
was first proposed in ~\cite{deflation-eurosys19} but required OS and application cooperation, while we focus on a hypervisor-only deflation approach.

\noindent \textbf{Vertical scaling with performance differentiation} for a single server under resource pressure due to increasing application load and server overbooking has been well studied in the past~\cite{lakew2015performance, Padala:2009, rao2013qos}. All previous work we are aware of tackles the problem 
of performance differentiation for a single server. Our work focuses on cluster-wide performance optimization when resources are deflated across the whole cluster.
Application performance models and workload prediction is a key component of elastic scaling~\cite{gong2010press, nguyen2013agile, padala2007adaptive, shen2011cloudscale,ali2012adaptive}. 
In contrast, deflation is a black-box, application agnostic, and reactive technique for handling resource pressure. 
Our deflatable VMs use a combination of overcommitment mechanisms that are adapt to application resource usage, and we consider the simultaneous deflation of \emph{all} resources. 
Deflation also exposes an explicit performance tradeoff, whereas elastic scaling approaches typically only reclaim unused resources.

\vspace*{-4mm}
\section{Conclusions}
\vspace*{-5pt}
\label{sec:conclusions}

In this paper we proposed the notion of deflatable VMs for running low-priority interactive applications. 
Deflatable VMs allow applications to continue running on transient resources, while minimizing the risk of preemptions and the associated downtimes. 
Our VM deflation mechanisms and cluster-level deflation policies reduce the performance overhead of applications and allow cloud platforms to increase cluster overcommitment and revenue.
The performance of deflatable VMs is within 10\% of their undeflated allocation---making them a viable alternative to current cloud transient VMs.

\noindent \textbf{Acknowledgments.} We wish to thank all the anonymous reviewers and our shepherd Renato Figueiredo, for  their insightful comments and feedback.
This research was supported by NSF grants 1836752, 1763834, and 1802523. 

\vspace*{-7pt}

{
\footnotesize
\bibliographystyle{acm} 
\bibliography{sample}
}

\end{document}